\documentclass[aps,prd,twocolumn,showpacs,superscriptaddress,groupedaddress,nofootinbib]{revtex4-1}  

\usepackage{amsmath}
\usepackage{amsfonts}
\usepackage{amssymb}
\usepackage{braket}
\usepackage[separate-uncertainty=true]{siunitx}

\usepackage{url}
\usepackage[colorlinks=true, urlcolor=blue, citecolor=blue, linkcolor=blue, filecolor=blue]{hyperref}

\usepackage{graphicx}     
\usepackage{multirow}
\usepackage{booktabs}

\DeclareDocumentCommand\d{}{\operatorname{d}\!}

\begin{document}

\title{Spectral, statistical and vertex functions in scalar quantum field theory\\ far from equilibrium}

\author{Linda Shen}
\author{J\"urgen Berges}
\affiliation{Heidelberg University, Institute for Theoretical Physics, Philosophenweg 16, 69120 Heidelberg, Germany}

\date{\today}

\begin{abstract}
We compute the far-from-equilibrium dynamics of relativistic scalar quantum fields in 3+1 space-time dimensions starting from over-occupied initial conditions. We determine universal scaling exponents and functions for two-point correlators and the four-vertex in a self-similar regime in time and space or momenta. The scaling form of the momentum-dependent four-vertex exhibits a dramatic fall-off towards low momenta. Comparing spectral functions (commutators) and statistical correlations (anti-commutators) of field operators allows us to detect strong violations of the fluctuation-dissipation relation in this non-perturbative infrared regime. Based on a self-consistent expansion in the number of field components to next-to-leading order, a wide range of interaction strengths is analyzed and compared to weak-coupling estimates in effective kinetic theory and classical-statistical field theory. 
\end{abstract}

\maketitle

\section{Introduction}

Non-equilibrium dynamics of relativistic scalar quantum field theory is an important cornerstone in our understanding of the evolution of the early universe. For a significant class of inflationary scenarios, where the scalar field describes the inflaton or even the Higgs field, the evolution traverses a far-from-equilibrium regime triggered by dynamical instabilities~\cite{Traschen:1990sw,Kofman:1994rk}. In general, these processes create highly occupied excitations at some characteristic momentum scale $Q_0$, which differs significantly from the temperature of a thermally equilibrated system with the same energy density and particle number. The over-occupied system subsequently relaxes in terms of self-similar cascades, transporting particles towards low momentum scales~\cite{Berges:2008wm,Orioli:2015dxa} and energy to higher momenta~\cite{Micha:2002ey,Micha:2004bv}. Similar phenomena are predicted to characterize the dynamics of the highly excited quark gluon plasma during the early stages of a heavy-ion collision~\cite{Schlichting:2012es,Kurkela:2012hp,Berges:2013eia,Berges:2013fga}. Recently, self-similar scaling phenomena have been experimentally discovered in ultra-cold quantum gases far from equilibrium, showing remarkable universal properties in the non-perturbative infrared regime at sufficiently low momenta~\cite{Prufer:2018hto,Erne:2018gmz,Prufer:2019kak}. Theoretical descriptions of the underlying non-thermal infrared fixed points typically involve effective kinetic theory~\cite{Berges:2010ez,Orioli:2015dxa,Chantesana:2018qsb,Walz:2017ffj,Mikheev:2018adp,Bhattacharyya:2019ffv} or classical-statistical approximations~\cite{Orioli:2015dxa,Moore:2015adu,Karl:2016wko,Mace:2016svc,Berges:2017igc,Deng:2018xsk,Berges:2019oun} in the weak coupling limit.   

In this work, we compute the time evolution of relativistic scalar quantum fields in 3+1 space-time dimensions starting from over-occupied initial conditions. We consider a self-interacting $N$-component field theory often employed in the context of scalar inflaton models, and which can also be taken to describe the Higgs sector of the standard model of particle physics for $N=4$. A self-consistent expansion in powers of $1/N$ to next-to-leading order (NLO) provides a non-perturbative account of the dynamics, such that we may analyze the highly occupied infrared for a wide range of interaction strengths. This has been previously employed to study far-from-equilibrium dynamics of this model, focusing on the role of a symmetry breaking field expectation value for its evolution~\cite{Berges:2016nru}. Here, we analyze for the first time the self-similar scaling properties of two-point correlators and the four-vertex in quantum field theory at NLO without further approximations by numerically extracting the universal dynamical exponents and scaling functions. Our results are compared to previous weak-coupling estimates for {\em equal-time} two-point correlators using effective kinetic theory or classical-statistical field theory~\cite{Orioli:2015dxa}. 

A further important focus of our work is the computation of {\em unequal-time} correlation functions far from equilibrium, namely the commutator expectation value of two field operators and the respective anti-commutator at different times. The former gives the spectral function, which provides essential information about the nature of the excitations, such as the possible existence or absence of long-lived quasi-particles far from equilibrium. In contrast, the anti-commutator expectation value captures quantum-statistical aspects, such as the occupation number of modes. While in thermal equilibrium the spectral (commutator) and statistical (anti-commutator) correlation functions are related by the fluctuation-dissipation relation, this is violated out of equilibrium in general. The question of whether some generalized fluctuation-dissipation relation may be defined, such as underlying effective kinetic descriptions of the dynamics, has been recently addressed in this context with the help of classical-statistical simulations~\cite{PineiroOrioli:2018hst,Boguslavski:2019ecc}. Here we compute both spectral and statistical correlation functions directly based on the underlying quantum field. While at sufficiently high momenta we recover the expected quasi-particle structure with a generalized fluctuation-dissipation relation, we demonstrate that significant violations occur in the non-perturbative infrared regime.    

The paper is organized as follows: In Sec.~\ref{sec:scalarQFT} we introduce the model and specify the class of far-from-equilibrium initial conditions we consider. The results for the equal-time two-point correlation functions and the four-vertex are presented in Sec.~\ref{section:self-similar_dynamics}. We extract the self-similar scaling properties for a wide range of couplings, and compare the full numerical results to analytical estimates that are obtained assuming universal scaling. In Sec.~\ref{sec:unequal_times} also unequal-time correlations are computed and the spectral as well as statistical functions are analyzed. We identify three characteristic momentum regimes and investigate the role of the fluctuation-dissipation relation in each of these regimes. We conclude in Sec.~\ref{sec:conclusion}. Three appendices provide details about the error estimates for the extraction of the scaling exponents, the Wigner transformation employed in the read-out of the spectral function, and the procedure for the numerical computation of the time evolution.

\section{Scalar quantum field theory far from equilibrium}
\label{sec:scalarQFT}

We consider a relativistic $N$-component scalar quantum field theory interacting via a quartic self-coupling $ \lambda $. Its $ O(N) $-symmetric classical action for massless fields is given by
\begin{align}\label{eq:action}
S &= \int  \!\mathrm{d}^4 x  \left[   \dfrac{1}{2} \partial _\mu  \varphi_a(x) \partial ^\mu  \varphi_a(x) 
- \dfrac{\lambda}{4!N}  \left(\varphi_a(x) \varphi_a(x)\right )^2 \right], 
\end{align}
where a summation over field indices $ a = 1, \ldots , N $ and Lorentz indices $ \mu  = 0, 1, 2, 3 $ is implied, with four-vector $x= (x^0,\mathbf{x})$. 

Quantum corrections are taken into account using a large-$N$ expansion at next-to-leading order (NLO) of the two-particle irreducible (2PI) effective action~\cite{Berges:2001fi,Aarts:2002dj}. This self-consistent expansion scheme is uniform in time, resumming secular terms such that it can be applied also to study late-time dynamics~\cite{Berges:2004yj}. 

A full description of general out-of-equilibrium dynamics can be based on both commutator and anti-commutator expectation values of products of field operators. Here we consider the statistical ($F$) and the spectral two-point function ($\rho$),
\begin{subequations}
\begin{align}
	F_{ab }(x,y) 
	&= \dfrac{1}{2} \left\langle\left\{ \hat{\varphi}_a(x),\hat{\varphi}_b(y)\right\}\right\rangle -  \left\langle\hat{\varphi}_a(x)\right\rangle \left\langle\hat{\varphi}_b(y)\right\rangle, 
	\label{eq:boson_statistical_fct}\\
	\rho _{ab}(x,y) 
	&= i \left\langle \left[ \hat{\varphi}_a(x),\hat{\varphi}_b(y)\right]\right\rangle,
	\label{eq:boson_spectral_fct}
	\end{align}%
	\label{eq:decomposition_F_R}
\end{subequations}
where $\{.\, ,.\, \}$ describes the anti-commutator and $[.\,,.\,]$ the commutator that is applied to the field operator $\hat{\varphi}_a(x)$.

The evolution equations at NLO in a $1/N$ expansion have been derived in Ref.~\cite{Berges:2001fi}. For their solution we have to supply initial conditions at time $x^0 = t = 0$. Here we employ initial conditions for the spatially homogeneous system with a vanishing macroscopic field, 
i.e.\ \mbox{$ \langle\hat{\varphi}_a\rangle (t=0) = 0 $} and $ \langle\partial_t\hat{\varphi}_a\rangle (t=0) = 0 $ for all field components.
By virtue of the $ O(N) $ symmetry, the field expectation value then remains zero at all times. This allows us to write 
\begin{equation}
F_{ab}(x,y)= F(x,y) \delta_{ab} \, , \quad \rho_{ab}(x,y)= \rho(x,y) \delta_{ab},
\end{equation} 
such that we only need to consider the diagonal elements $ F(x,y) $ and $ \rho (x,y)$. 

The initial conditions at $ t = 0 $ for the spectral function are fixed by its anti-symmetry, $\rho (t,t, \mathbf{x}, \mathbf{y})=0$, and the equal-time commutation relations of the bosonic quantum theory, 
\begin{align}
\partial _t \rho (t,t' , \mathbf{x}, \mathbf{y}) \Big|_{t=t'}&= \delta (\mathbf{x-y}). \label{eq:commrel}
\end{align}
For the statistical function at initial time we consider Gaussian correlations, which in spatial Fourier space with momentum $\mathbf{p}$ are parameterized as 
\begin{align}
F (t, t', \mathbf{p})\Big|_{t=t'=0}&= \dfrac{f_\mathbf{p} + 1/2}{\omega_\mathbf{p}} \cos \left[
\omega_\mathbf{p} (t-t') \right]\Big|_{t=t'=0}. 
\label{eq:free_solution_F}
\end{align} 
Here $ f_\mathbf{p} $ denotes an initial particle number distribution with dispersion $ \omega_\mathbf{p} $. 
The latter is set at initial time by $ \omega_\mathbf{p} =  \sqrt{  |\mathbf{p}|^2  + m^2_0} $ in the limit of a vanishing initial mass $ m_0 \rightarrow 0^+$. Following Ref.~\cite{Berges:2016nru}, we consider the initial condition of a highly occupied system of particles with distribution function
\begin{align}
f_\mathbf{p} = \dfrac{n_0}{\lambda} \, \Theta(Q_0 - |\mathbf{p}| ) \,,
\label{eq:initial_distribution}
\end{align}
where the characteristic momentum $ Q_0 $ sets the initial scale, with occupancy parameter $ n_0 $, and Heaviside function $\Theta$. Since the initial occupancy is inversely proportional to the coupling, this represents a non-perturbative problem even for $\lambda \ll 1$. Since the NLO approximation for the dynamics is non-Gaussian, higher correlations will build up during the time evolution.

Starting from this over-occupied initial situation, we compute the time evolution of the system by numerically solving the coupled NLO evolution equations for the statistical and spectral correlation functions~\cite{Berges:2001fi}:
\begin{subequations}
\begin{align}
\left[ \Box_x + M^2(x)\right] F(x,y) 
&=
- \int_{t_0}^{x^0}{\!\mathrm{d} z\, \Sigma_\rho(x,z) F(z,y)} 		
\nonumber\\
& \quad +\int_{t_0}^{y^0}{\!\mathrm{d} z\, \Sigma_F(x,z) \rho(z,y)}\,,	\\
\left[\Box_x  + M^2(x)\right] \rho(x,y)
&=
- \int_{y^0}^{x^0}{\!\mathrm{d} z\, \Sigma_\rho(x,z) \rho(z,y)}\,,
\end{align}\label{eq:EoM}%
\end{subequations}
with shorthand notation $\int_{t_1}^{t_2}{\!\d z} \equiv \int_{t_1}^{t_2}{\!\d z^0 \int{\!\d^3 z}}$ and the effective mass squared $  M^2(x) = \lambda\,  \frac{N+2}{6N} F(x,x)$.
The NLO approximation for the statistical and spectral components of the self-energies, $\Sigma_F(x,z)$ and $\Sigma_\rho(x,z)$, entails a geometric series summation of the correlation functions $F(x,y)$ and $\rho(x,y)$, which can be conveniently expressed in terms of the summation functions $I_F(x,y)$ and $I_\rho(x,y)$: 
\begin{subequations}
\begin{align}
\Sigma_F(x,y) 
&=
-\frac{\lambda}{3N} \Big( F(x,y) I_F(x,y) - \frac{1}{4} \rho(x,y) I_\rho(x,y) \Big) \,, 
\\
\Sigma_\rho(x,y) 
&=
-\frac{\lambda}{3N} \Big( F(x,y) I_\rho(x,y) + \rho(x,y) I_F(x,y) \Big) \,.
\end{align}
\label{eq:sigmas}%
\end{subequations}
The summation functions
\begin{subequations}
\begin{align}
I_F(x,y) = \Pi_F(x,y) &- \int_{t_0}^{x^0}{\!\!\!\!\d z\, I_\rho(x,z) \Pi_F(z,y)} \nonumber\\
&+ \int_{t_0}^{y^0}{\!\!\!\!\d z\, I_F(x,z) \Pi_\rho(z,y)} \,,
\\
I_\rho(x,y) = \Pi_\rho(x,y) &- \int_{y_0}^{x^0}{\!\!\!\!\d z\, I_\rho(x,z) \Pi_\rho(z,y)} \,,
\end{align}\label{eq:resummation_functions}%
\end{subequations}
contain as building blocks the `one-loop' self-energies $\Pi_F(x,y)$ and $\Pi_\rho(x,y)$ with 
\begin{subequations}
\begin{align}
\Pi_F(x,y)  &= \frac{\lambda}{6} \left( F^2(x,y) - \frac{1}{4} \rho^2(x,y) \right) \, ,\\
\Pi_\rho(x,y) &= \frac{\lambda}{3}\, F(x,y) \rho(x,y) \, .
\end{align}%
\label{eq:oneloopself}%
\end{subequations}
The latter spectral component is directly related to the retarded self-energy 
\begin{align}
\Pi_R(x,y) &=\Theta(x^0 - y^0) \Pi_\rho(x,y)\,.
\label{eq:oneloop_ret_adv}%
\end{align}

We emphasize that the large $ N $ approximation only involves a systematic power counting of factors of $1/N$. As a consequence, the terms neglected are suppressed by an additional power of $1/N$. 
Since this is not an expansion in powers of the coupling $\lambda$, even non-perturbative situations far from equilibrium such as the over-occupied initial conditions (\ref{eq:initial_distribution}) can be addressed. Moreover, the NLO contributions are essential: at LO ($N \rightarrow \infty$) the right hand side of the evolution equations (\ref{eq:EoM}) would vanish since the self-energies (\ref{eq:sigmas}) are zero at that order. 

The numerical results presented in this work employ $ N=4 $ and occupation parameter $ n_0 =100$. We study a wide range of coupling parameters $ \lambda = 0.01, 0.10, 1.0, 2.0 $. If not stated otherwise, $ \lambda=1.0 $ is used. In the following, all quantities are given in units of the characteristic initial scale $ Q_0 $ and are stated as dimensionless numbers. The equations of motion are discretized on a lattice with temporal step $a_t$ and spatial spacing $a_s$. The results shown are obtained from computations using $ N_s = 500$ points on a spatial grid with $ a_s  = 0.75 $, corresponding to an ultraviolet (UV) momentum cutoff of $  \Lambda_{\mathrm{UV}} = 4.19$ and an infrared (IR) cutoff of $ \Lambda_{\mathrm{IR}} = 0.0084 $. 
We have checked that all relevant results are insensitive to both IR and UV cutoffs. 
In order to efficiently approach late times, the time step $a_t$ is tuned to be as large as possible while checking numerical convergence to runs with smaller time steps. 
For the presented numerical results a time step of $ a_t  = 0.3$ is used. 

\section{Universal scaling dynamics of equal-time correlations}
\label{section:self-similar_dynamics}

We first compute the non-equilibrium evolution of the statistical correlation function (\ref{eq:boson_statistical_fct}) at equal times, for which the spectral function (\ref{eq:boson_spectral_fct}) vanishes because of its anti-symmetry. A similar calculation for a larger class of initial conditions (with non-zero initial field expectation value) has been done in Ref.~\cite{Berges:2016nru} pointing out the independence of rescaled results on system parameters, such as the value of the coupling $\lambda$ or details of the initial conditions, in an emergent universal regime associated to a non-thermal fixed point~\cite{Berges:2008wm,Orioli:2015dxa}. Exploiting this insensitivity to initial condition details, we restrict ourselves to the symmetric regime with initial conditions described in Sec.~\ref{sec:scalarQFT}. This reduces the numerical efforts and we do not have to employ an adaptive grid size as done in Ref.~\cite{Berges:2016nru}, which allows us for the first time to accurately extract the universal self-similar scaling exponents and scaling function from the complete NLO evolution equations (\ref{eq:EoM}) -- (\ref{eq:oneloopself}) by numerical computations. A calculation of the far-from-equilibrium scaling exponents and function has so far only been done using additional assumptions, such as a quasi-particle ansatz for an effective kinetic description at the non-thermal fixed point in Refs.~\cite{Orioli:2015dxa,Walz:2017ffj,Chantesana:2018qsb}, or based on classical-statistical field theory approximations in Ref.~\cite{Orioli:2015dxa} in the weak-coupling limit. We emphasize that our approach is not restricted to weak couplings and includes genuine quantum effects at NLO in the large-$N$ expansion.

\subsection{Particle distribution}
\label{section:scaling_occ}

\begin{figure} 
\centering
\includegraphics[width=.5\textwidth]{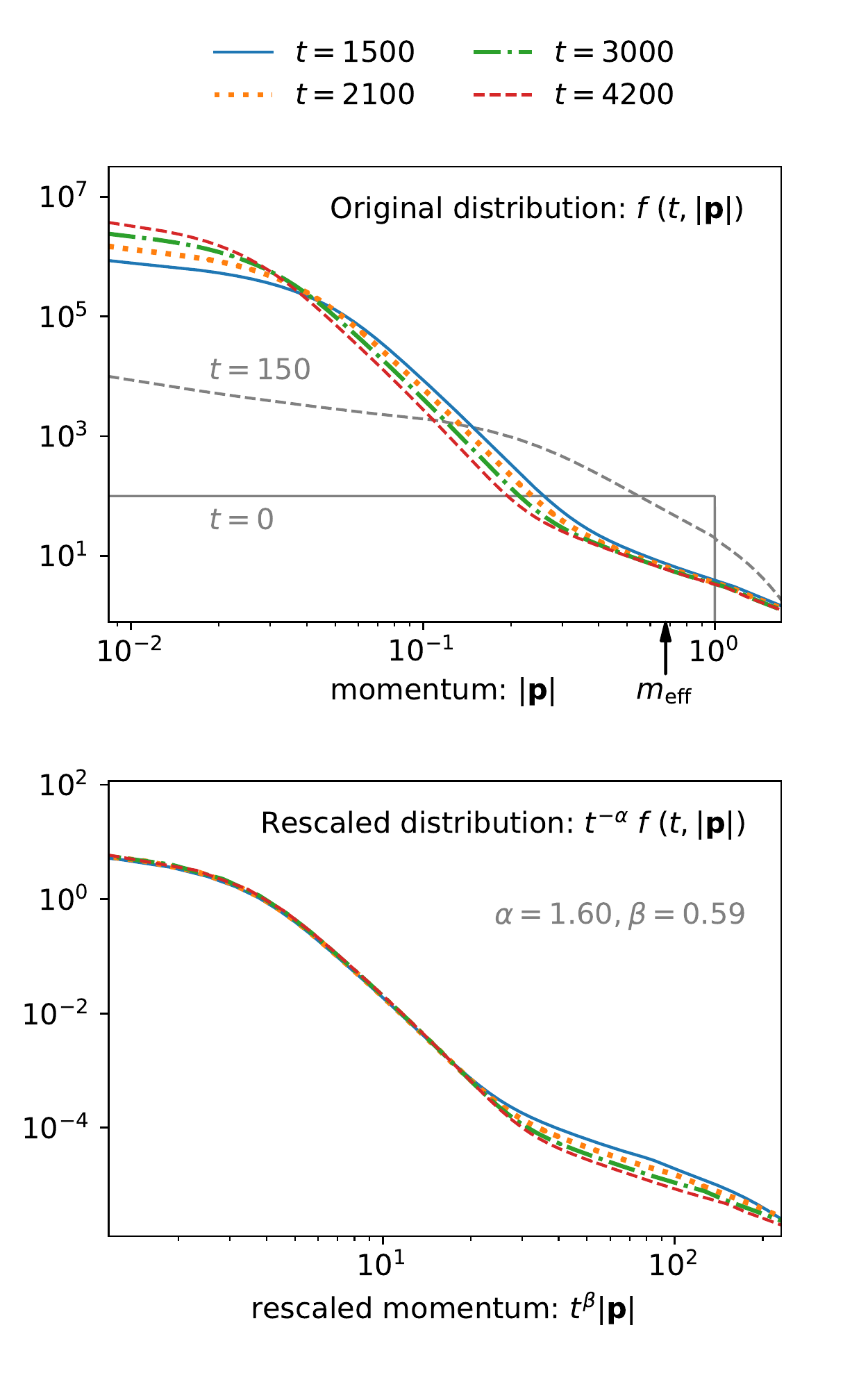}	
\caption{
	The upper graph shows the distribution function as a function of momentum at different times. 
	In the lower plot, the rescaled distribution versus rescaled momentum is given. 
}
\label{fig:Occupation_Original_and_Rescaled}
\end{figure}

\begin{figure*}
	\centering
	\includegraphics[width=1.0\textwidth]{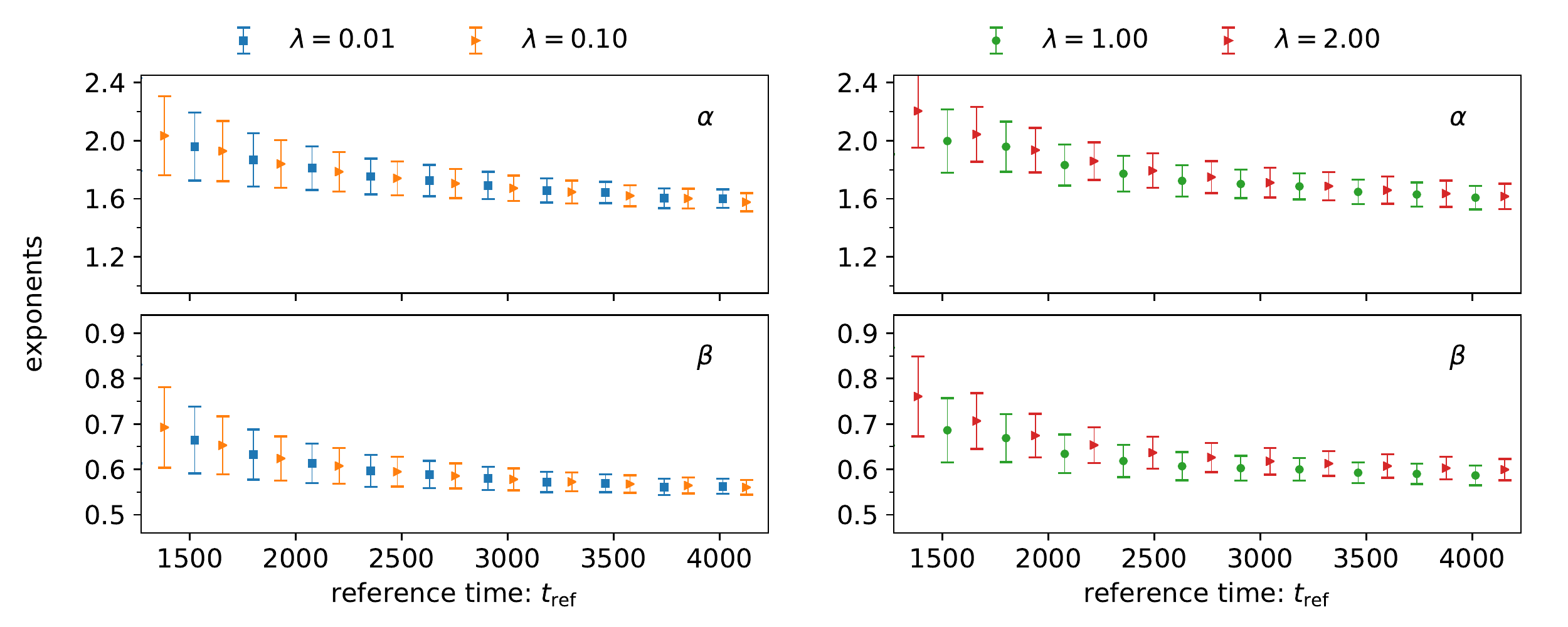}	
	\caption{
		The scaling exponents $ \alpha $ and $\beta$ for the self-similar behavior of the distribution function $ f(t, |\mathbf{p}|) $ extracted at reference times $ t_\mathrm{ref} $, shown for different values of the coupling parameter $ \lambda $.  
	}
	\label{fig:scaling_exponents}
\end{figure*}

So far, the phenomenon of scaling has mainly been discussed in terms of a particle number distribution function, whose time-dependence we extract from the two-point correlation function as~\cite{Berges:2001fi,Berges:2004yj}
\begin{equation}
f(t, |\mathbf{p}| )  + \dfrac{1}{2}
= \sqrt{F(t,t', |\mathbf{p}|)\partial _t \partial _{t'} F(t,t', |\mathbf{p}|)} \ \Big | _{t=t'} \label{eq:equaltime_particle_distribution} \, .
\end{equation}
Similarly, one may define a time-dependent effective dispersion
\begin{equation}
\omega(t, |\mathbf{p}| )  
= \sqrt{\dfrac{\partial _t \partial _{t'} F(t,t', |\mathbf{p}|)}{F(t,t', |\mathbf{p}|)}} \ \Bigg | _{t=t'}, 
\label{eq:equaltime_dispersion}
\end{equation}
such that we have at initial time $ f(t=0, |\mathbf{p}|) = f_\mathbf{p}$ and $ \omega(t=0, |\mathbf{p}|) = \omega_\mathbf{p} $ according to (\ref{eq:free_solution_F}).

Starting from the over-occupied initial state at $ t=0 $, the non-equilibrium evolution leads to a redistribution of both particle number and mode energy. In the upper plot of Fig.~\ref{fig:Occupation_Original_and_Rescaled}, we show the time evolution of the distribution function $f(t, \mathbf{p})$. The effective mass $m_{\mathrm{eff}}$, which is given by the approximately time-independent value of the dispersion at zero momentum as analyzed in Sec.~\ref{sec:effmass}, is also indicated. 
For $ t \gtrsim 1500 $, the dynamics slows down considerably and we analyze in the following whether the system becomes self-similar. We concentrate on the non-perturbative behavior for sufficiently low momenta, and refer for the analysis of the perturbative high-momentum properties to Refs.~\cite{Micha:2004bv,Berges:2016nru}.

For a self-similar time evolution the distribution obeys the scaling property
\begin{align}
f(t, \mathbf{p}) = t^{\alpha} f_S(t^{\beta}|\mathbf{p}|)\,, 
\label{eq:scaling_function}
\end{align}
with scaling exponents $\alpha$, $\beta$ and time-independent scaling function $ f_S $. Therefore, in the scaling regime $ t^{-\alpha} f(t, |\mathbf{p}|) $ does not depend on time and momentum separately but only on the product $ t^\beta |\mathbf{p}| $ for a set of exponents $ \alpha$ and $\beta $. Universality implies that the shape of the distribution function and the values of the exponents do not depend on the microscopic model parameters, such as the value of the coupling $\lambda$, which is discussed in the following.

As shown in the lower plot of Fig.~\ref{fig:Occupation_Original_and_Rescaled}, the distribution function at different times can be rescaled such that the curves of $ t^{-\alpha} f(t, |\mathbf{p}|) $ as a function of $ t^\beta |\mathbf{p}| $ lie on top of each other for lower momenta. 

Numerically, the scaling exponents are obtained by comparing the distribution function at some reference time $ t_\mathrm{ref} $ with several earlier times $ t $, where we perform comparisons within a time window of $ \Delta t_\mathrm{w} = 720$. For details on the method and the employed error estimation we refer to Appendix~\ref{appendix:scaling_exponents_method}. In Fig.~\ref{fig:scaling_exponents} we show our results for the scaling exponents $ \alpha $ and $ \beta $ for different couplings where the data points are binned for the plots. 
Both exponents approach approximately constant values given in Table~\ref{tab:exponent_values}. 
For later comparison, we also present values for the exponent $ \alpha_\lambda $ as defined in the table caption. 

The exponents for all couplings studied here agree within errors with each other. Furthermore, they are consistent with the results found in Ref.~\cite{Orioli:2015dxa} using classical-statistical lattice simulations, confirming that statistical fluctuations dominate over genuine quantum fluctuations in this highly occupied regime. As pointed out in Refs.~\cite{Orioli:2015dxa,Chantesana:2018qsb,Walz:2017ffj}, the values of these exponents coincide with those of the corresponding non-relativistic scalar model, since infrared momenta below $m_{\mathrm{eff}}$ of the relativistic theory behave non-relativistically. Within errors, we also find $\alpha = d \beta$ for $d=3$ spatial dimensions such that $\int_\mathbf{p} f(t,|\mathbf{p}|) = t^{\alpha - d \beta} \int_\mathbf{q} f_S(|\mathbf{q}|)$ is approximately conserved, reflecting a transport of particles towards lower momenta for the $\beta > 0$ observed. 

\begin{table}[h]
	\centering
	\begin{ruledtabular}
		\begin{tabular}{cccccc}
		&$\lambda$& $\alpha$ & $\beta$& $2 ( \beta - \alpha )$\\
			\hline
&$\SI{0.01}{} $ & \SI{1.59(6)}{}& \SI{0.56(2)}{}& \SI{-2,06(13)}{}&\\
&$\SI{0.10}{} $ & \SI{1.60(6)}{}& \SI{0.57(2)}{}& \SI{-2,06(13)}{}&\\
&$\SI{1.00}{} $ & \SI{1.60(8)}{}& \SI{0.59(2)}{}& \SI{-2,02(17)}{}&\\
&$\SI{2.00}{} $ & \SI{1.61(9)}{}& \SI{0.60(2)}{}& \SI{-2,02(19)}{}&\\
		\end{tabular}
	\end{ruledtabular}
	\caption{Exponents $\alpha$ and $\beta$ obtained from the scaling analysis of the particle distribution $ f(t, |\mathbf{p}|) $. 
	Using these values we also give $ \alpha_\lambda = 2 ( \beta - \alpha ) $ for later comparison, which is the expected scaling exponent of the four-vertex 
	as introduced in Sec.~\ref{sec:fourvertex}. 
	Our results are shown for different values of the coupling parameter $ \lambda $.}
\label{tab:exponent_values}
\end{table}

\subsection{Mode energy}

\begin{figure} 
	\centering
	\includegraphics
	[width=.5\textwidth]
	{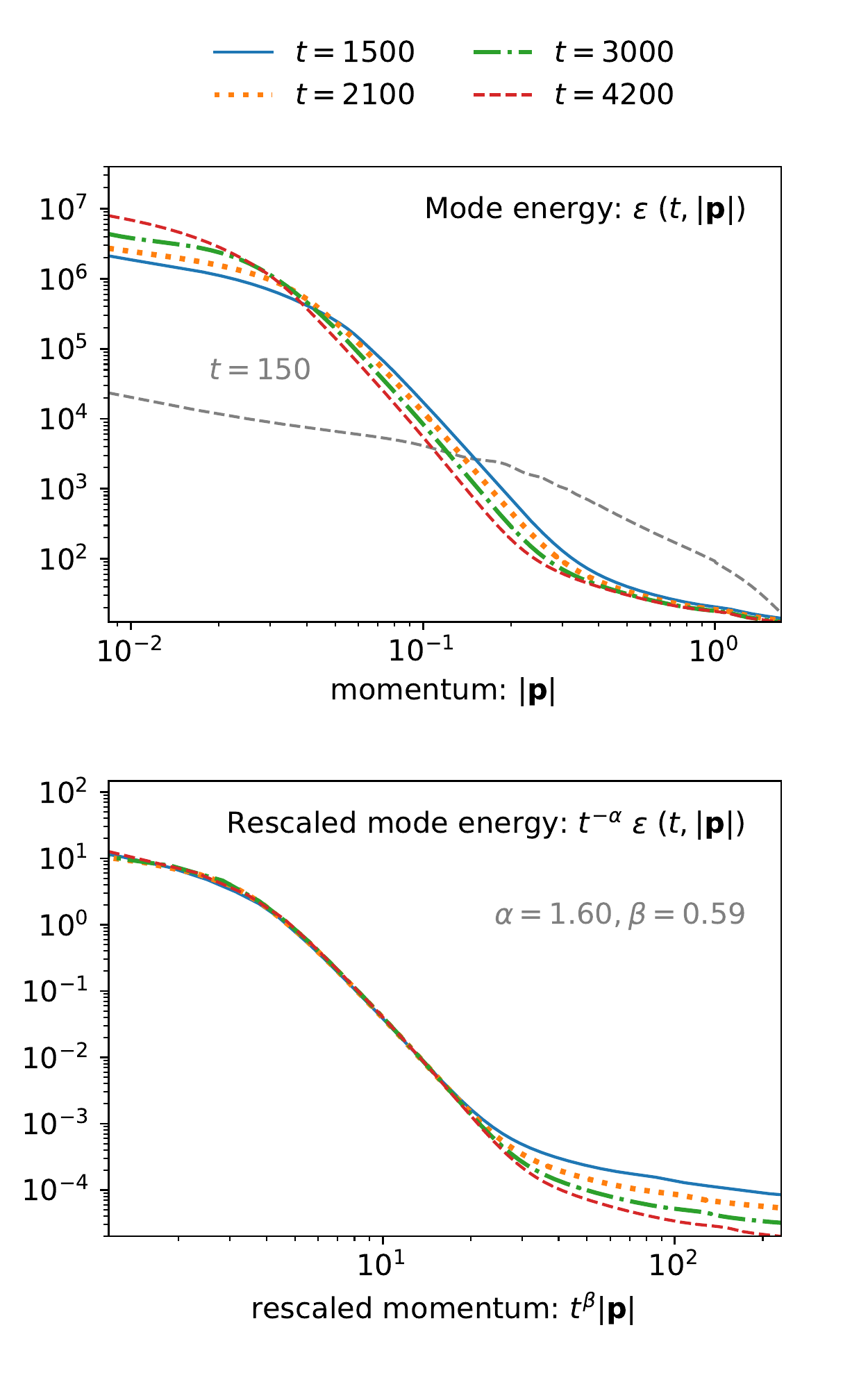}	
	\caption{
		The original and rescaled mode energies at different times. The same exponents as obtained from the scaling analysis of the distribution function $f$ are used. 
	}
	\label{fig:Energy_distribution_Original_and_Rescaled}
\end{figure}

\begin{figure} 
\centering
\includegraphics
	[width=.5\textwidth]
	{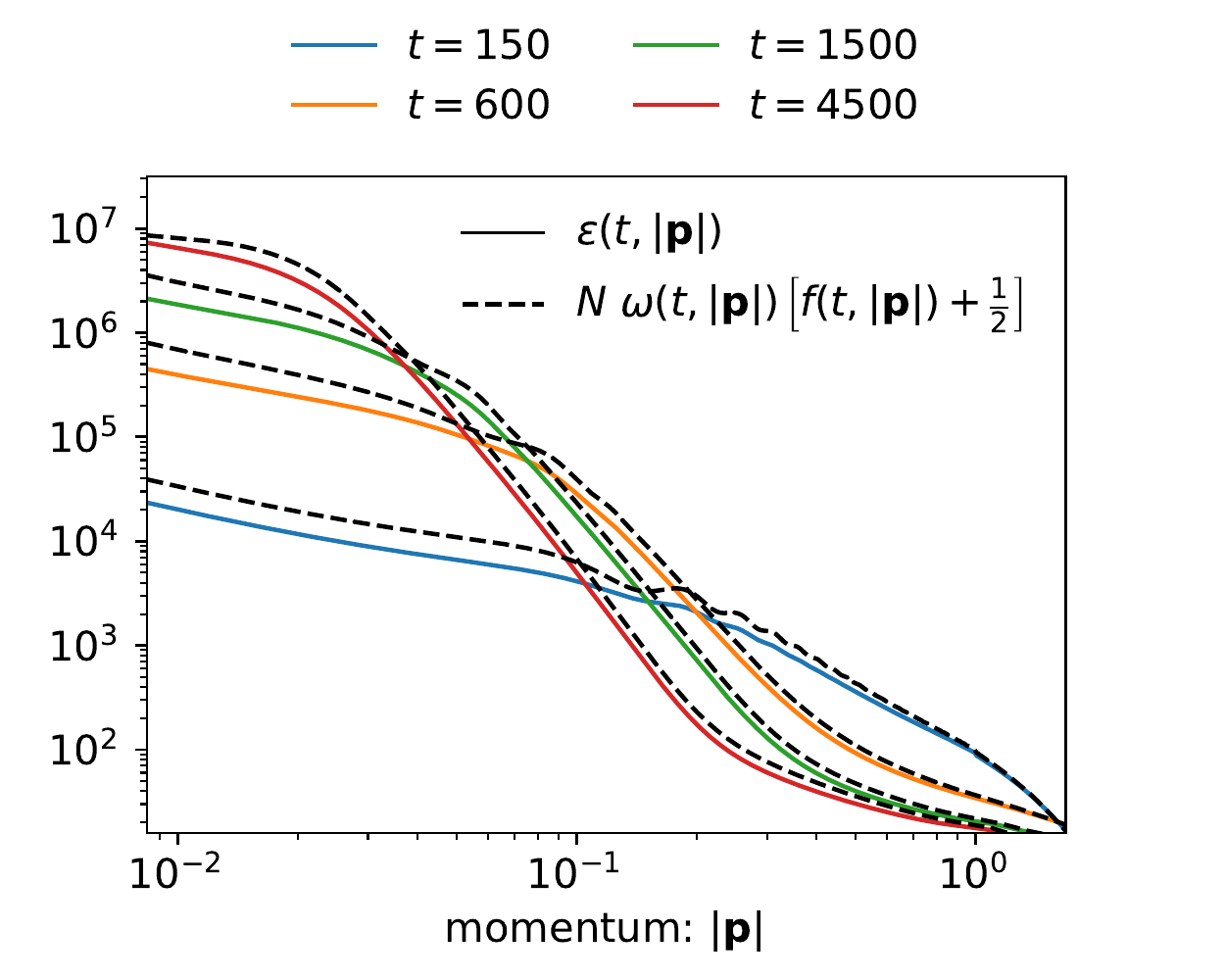}	
\caption{
	Comparison of the mode energy at NLO (\ref{eq:modeenergy}) and the quasi-particle expression (\ref{eq:appmodeenergy}) as a function of momentum for different times. 
}
\label{fig:modenergycomparison}
\end{figure}

In addition, we consider the time-dependent mode energy $\varepsilon(t,|\mathbf{p}|)$, which is given at NLO in the $1/N$ expansion by \cite{Berges:2016nru}
\begin{align}
\frac{\varepsilon(t,|\mathbf{p}|)}{N} 
&=
\left[ \frac{\partial_t\partial_{t'}}{2} + \frac{\mathbf{p}^2}{2} +\frac{\lambda}{4!} \int_\mathbf{q}\!\! F(t,t,|\mathbf{q}|) \right]\! F(t,t',|\mathbf{p}|)\Big|_{t= t'}
\nonumber\\
&\qquad + \frac{1}{2N} \, I_F(t,t,|\mathbf{p}|) \label{eq:modeenergy}\\[1em]
&\simeq \, \omega(t,\mathbf{p})\, \left[f(t,|\mathbf{p}|) + \frac{1}{2}\right]\, ,
\label{eq:appmodeenergy}
\end{align}
where the approximation employed for the last line is only used here to analyze the quasi-particle content of the dynamics.\footnote{For a proper quasi-particle description of the energy density one may subtract the zero-mode quantum-half appearing in (\ref{eq:appmodeenergy}), which we do not consider here for the lattice regularized theory.}
The momentum sum of the mode energy (\ref{eq:modeenergy}) is equal to the conserved total energy density at NLO, $ \varepsilon(t) = \int \mathrm{d}^3 \mathbf{p}/ (2\pi)^3  \varepsilon (t, |\mathbf{p}|) $. In our simulations, we checked that $ \varepsilon(t)$ is conserved at the level of $\SI{1}{\percent} $ accuracy for the times under consideration. 

The scaling analysis for the mode energy density employs
\begin{align}
\varepsilon (t, |\mathbf{p}|) = t^{\alpha}\, \varepsilon_S( t^{\beta}  |\mathbf{p}| ) \,, 
\end{align}
where we anticipate that the scaling exponents are the same as for the distribution function, and $ \varepsilon_S $ denotes the energy scaling function. The upper graph of Fig.~\ref{fig:Energy_distribution_Original_and_Rescaled} shows the time evolution of $\varepsilon (t, |\mathbf{p}|)$, while the lower one displays the rescaled quantity $t^{-\alpha} \varepsilon(t, |\mathbf{p}|)$ as
a function of $t^{\beta} |\mathbf{p}|$ employing the same values for the exponents as obtained from the particle number distribution. One observes that in the scaling regime all curves collapse to a single one in the infrared to very good accuracy.

In order to illustrate this agreement, we analyze the approximate quasi-particle expression \eqref{eq:appmodeenergy}. In Fig.~\ref{fig:modenergycomparison} the mode energy at NLO \eqref{eq:modeenergy} is compared to the quasi-particle expression \eqref{eq:appmodeenergy}. One observes rather good agreement, in particular in terms of the scaling properties. 

\subsection{Scaling of the effective coupling}
\label{sec:fourvertex}

\begin{figure}
	\centering
	\includegraphics
	[width=.5\textwidth]
	{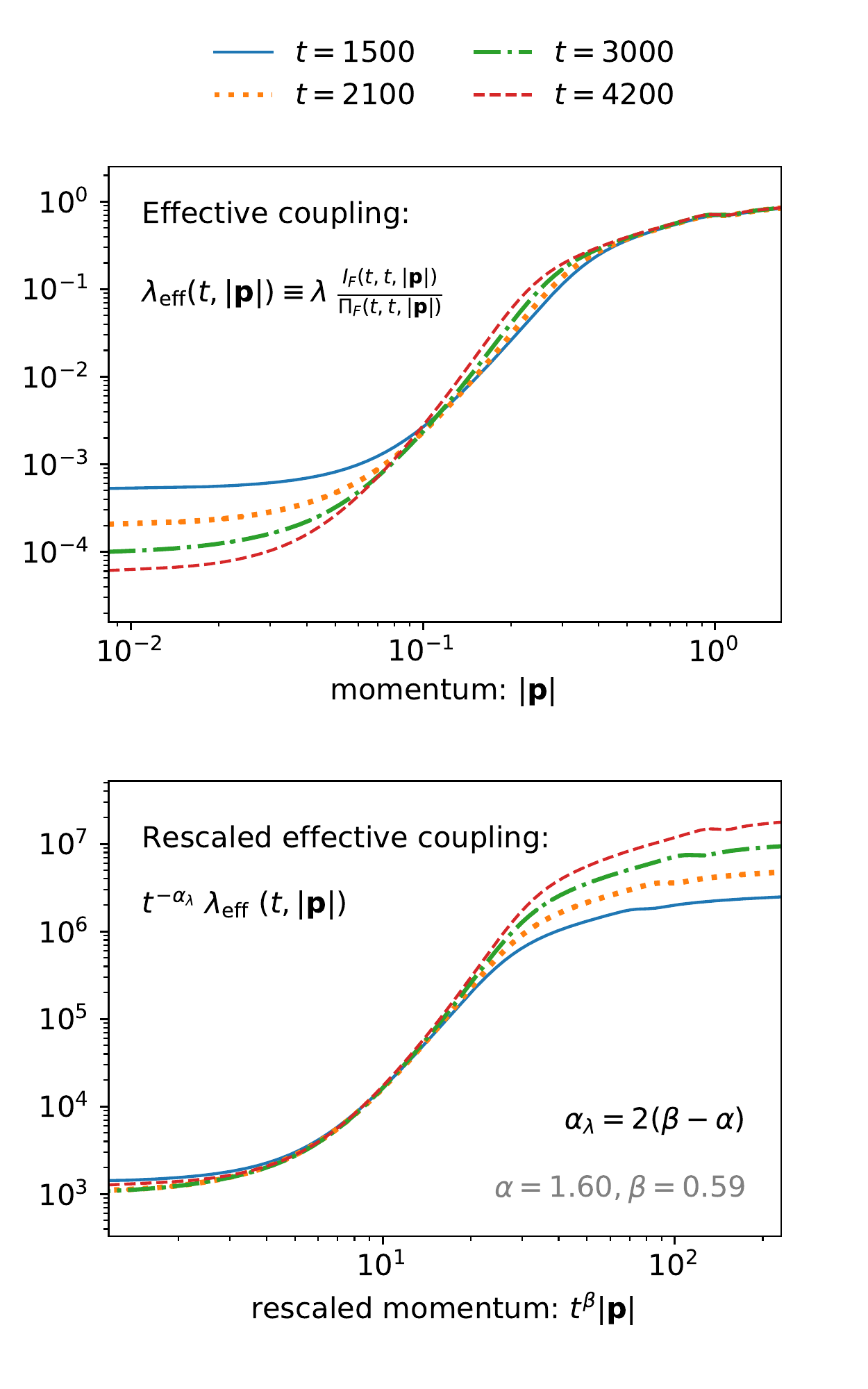}	
	\caption{
		The original and rescaled effective coupling for different times. 
		}
	\label{fig:SCA__Original_and_Rescaled_l_eff}
\end{figure}

In this section we analyze the scaling properties of the four-vertex in an approximation based on the large-$N$ expansion to NLO. Since we are interested in a slowly evolving self-similar scaling regime, we may simplify the
computation considerably by relying on a derivative expansion in time. More precisely, at lowest order in derivatives   
it is convenient to consider the (on-shell) effective coupling in Fourier space given by~\cite{berges2009nonthermal,Berges:2010ez,Orioli:2015dxa}
\begin{align}
\lambda_\mathrm{eff} (t, |\mathbf{p}|) = \dfrac{\lambda}{
	\left|		1 + \Pi_R(t, |\mathbf{p}|)	\right|^2
}\,.
\label{eq:eff_coupling1}
\end{align}
This effective coupling approximates the full four-vertex, which at this order receives its momentum dependence from the resummed geometric series underlying the NLO approximation~\cite{Berges:2001fi,Aarts:2002dj}. Moreover, the effective coupling is directly related to the summation functions (\ref{eq:resummation_functions}) according to~\cite{berges2009nonthermal,Berges:2010ez,Orioli:2015dxa} 
\begin{align}
\dfrac{\lambda_\mathrm{eff} (t, |\mathbf{p}|) }{\lambda}
	=  \dfrac{I_F(t, |\mathbf{p}|) }{\Pi_F(t, |\mathbf{p}|)} \,.
	\label{eq:eff_coupling2}%
\end{align}
In the following this is used to numerically compute the time-dependent effective coupling,
where $ I_F(t, |\mathbf{p}|) $ and $ \Pi_F(t,|\mathbf{p}|) $ are obtained from the expressions \eqref{eq:resummation_functions} and \eqref{eq:oneloopself} 
in spatial Fourier space evaluated at equal times. 

Using the scaling property (\ref{eq:scaling_function}) of the distribution function in the expression for the self-energies (\ref{eq:oneloopself}) and (\ref{eq:sigmas}) in the non-relativistic regime,
one expects~\cite{Orioli:2015dxa,Chantesana:2018qsb,Walz:2017ffj}  
\begin{align}
\lambda_\mathrm{eff}(t, |\mathbf{p}|) = t^{\alpha_\lambda} 
\lambda_{\mathrm{eff},S}(t^{\beta}|\mathbf{p}|)\,,
\label{eq:scaling_function_l_eff}
\end{align}
with scaling function $ \lambda_{\mathrm{eff},S} $ and coupling scaling exponent 
\begin{align}
\alpha_\lambda = -2\left[ (2-d) \beta + \alpha \right] \, 
\label{eq:exponents_relation}
\end{align}
directly related to the occupation number exponents $ \alpha  $ and $ \beta $ in $d$ spatial dimensions. 

The upper panel of Fig.~\ref{fig:SCA__Original_and_Rescaled_l_eff} shows the momentum dependence of the effective coupling at different times. Remarkably, the effective coupling drops over several orders of magnitude in the non-perturbative infrared regime, where the occupation number grows larger with time. This non-equilibrium phenomenon of a dynamically reduced four-vertex counteracts the dramatic Bose enhancement from the very high occupancies in the infrared, which would otherwise lead to faster and faster dynamics for growing occupancies. The corresponding phenomenon has recently also been experimentally demonstrated in an atomic Bose gas far from equilibrium~\cite{Prufer:2019kak}.

\begin{figure}
	\centering
	\includegraphics[width=.5\textwidth]{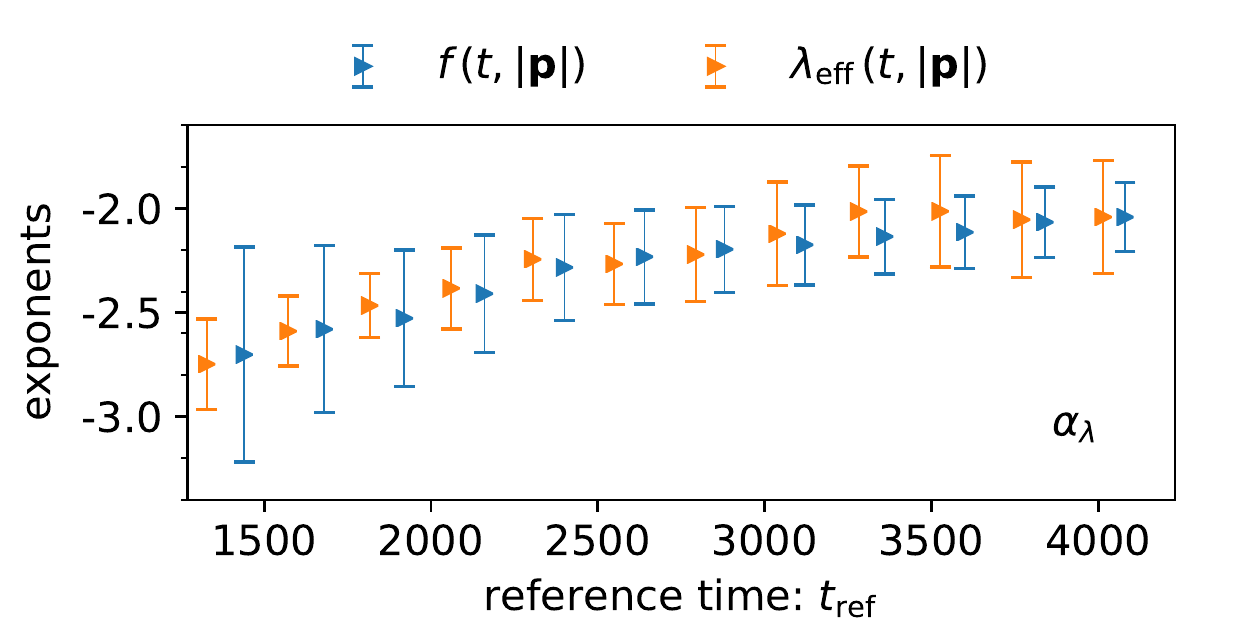}	
	\caption{
		The scaling exponents $ \alpha_\lambda $ obtained from the analysis of the distribution function $ f $ as well as the effective coupling $ \lambda_\mathrm{eff} $ for quartic self-interaction $ \lambda = 1.0 $. 
	}
	\label{fig:scaling_exponents_a_lambda}
\end{figure}

The lower graph of Fig.~\ref{fig:SCA__Original_and_Rescaled_l_eff} demonstrates that the numerical data for the effective coupling collapses rather well to a common curve in the infrared momentum range when rescaled accordingly. 
As a check, we also determine the scaling exponents $ \alpha_\lambda $ and $ \beta $ defined by \eqref{eq:scaling_function_l_eff} directly from our numerical data without assuming the scaling relation (\ref{eq:exponents_relation}), see Appendix \ref{appendix:scaling_exponents_method} for details on the method. 
The binned data obtained for $ \alpha_\lambda $ in this way is plotted in Fig.~\ref{fig:scaling_exponents_a_lambda}, where we also show the results for the exponent if computed according to \eqref{eq:exponents_relation} using the scaling exponents $\alpha$ and $\beta$. 
We observe a good agreement of the data although the errors for the analysis using $ \lambda_\mathrm{eff} $ are larger and fluctuate stronger, which is reflected in enhanced statistical errors as discussed in Appendix  \ref{appendix:scaling_exponents_method}. The asymptotic values approached are presented in Table~\ref{tab:exponent_values_l_eff}. 

\begin{table}[h]
	\centering
	\begin{ruledtabular}
		\begin{tabular}{cccccc}
			&$\lambda$& $\alpha_\lambda$ & $\beta$& $\beta - \alpha_\lambda / 2 $\\
			\hline
			&$\SI{0.01}{} $ & \SI{-2.00(42)}{}& \SI{0.60(2)}{}& \SI{1.60(25)}{}&\\
			&$\SI{1.00}{} $ & \SI{-2.01(42)}{}& \SI{0.67(2)}{}& \SI{1.68(25)}{}&\\
		\end{tabular}
	\end{ruledtabular}
	\caption{Scaling exponents for the relativistic scalar field theory. The exponents $ \alpha_\lambda $ and $\beta $ are obtained from the analysis of the effective coupling $ \lambda_\mathrm{eff} (t, |\mathbf{p}|) $. The last column shows the computed values for $ \alpha = \beta - \alpha_\lambda / 2    $.}
	\label{tab:exponent_values_l_eff}
\end{table}

\section{Spectral and statistical correlations at unequal times}
\label{sec:unequal_times}

For the study of correlation functions at different times $t$ and $t'$, it is convenient to rephrase the time-dependence in terms of Wigner coordinates, employing the central time $ \tau = (t + t')/2 $ and the relative time $ \Delta t = t-t' $. For the spatially homogeneous system, we then denote the two-point functions using the new temporal coordinates as $F(\tau, \Delta t, |\mathbf{p}|) $ and $\rho(\tau, \Delta t, |\mathbf{p}|) $. 

In order to study the frequency spectrum of these statistical and spectral functions, we consider a finite-range Fourier transformation of the Wigner space propagators with respect to the relative time $\Delta t $,
\begin{subequations}
\begin{align}
F (\tau, \omega, |\mathbf{p}|) 
&= \int  _{-2\tau}^{2\tau}\mathrm{d} \Delta t \ 
e^{i\omega \Delta t} 
F \left( \tau, \Delta t, |\mathbf{p}|\right ), 
\label{eq:FW_F}\\
i\tilde{\rho} (\tau, \omega, |\mathbf{p}|) 
&= \int  _{-2\tau}^{2\tau}\mathrm{d} \Delta t\ 
e^{i\omega \Delta t} 
\rho \left( \tau, \Delta t, |\mathbf{p}|\right ), 
\label{eq:FW_rho}
\end{align}
\label{eq:Wigner_transformation}%
\end{subequations} 
where the factor of $ i $ is introduced such that both $\tilde{\rho}(\tau, \omega, |\mathbf{p}|) $ and $ F(\tau, \omega, |\mathbf{p}|) $ are real. To ease the notation, we will neglect the tilde for the real spectral function in frequency space, $\tilde{\rho} (\tau, \omega, |\mathbf{p}|)$, in the following having in mind the extra factor of $i$ in its definition.  

The integrals with respect to the relative times  in (\ref{eq:Wigner_transformation}) are fundamentally restricted by $ \pm 2\tau$ for the initial value problems with $ t, t' \geq 0 $. Moreover, it is sufficient to present the propagators for positive $ \Delta t $ or $ \omega $, since the statistical (spectral) function is \mbox{(anti-)symmetric} in $ \Delta t $ and hence $ \omega $. Effects resulting from the finite-time boundary vanish in the limit $ \tau \rightarrow \infty $ and are discussed further in Sec.~\ref{section:fluct_diss_rel}. Details on the numerical treatment are presented in Appendix~\ref{appendix:Wigner_Fourier_trafo}.

\begin{figure}[t]
	\centering
	\includegraphics[width=0.5\textwidth, 
	]{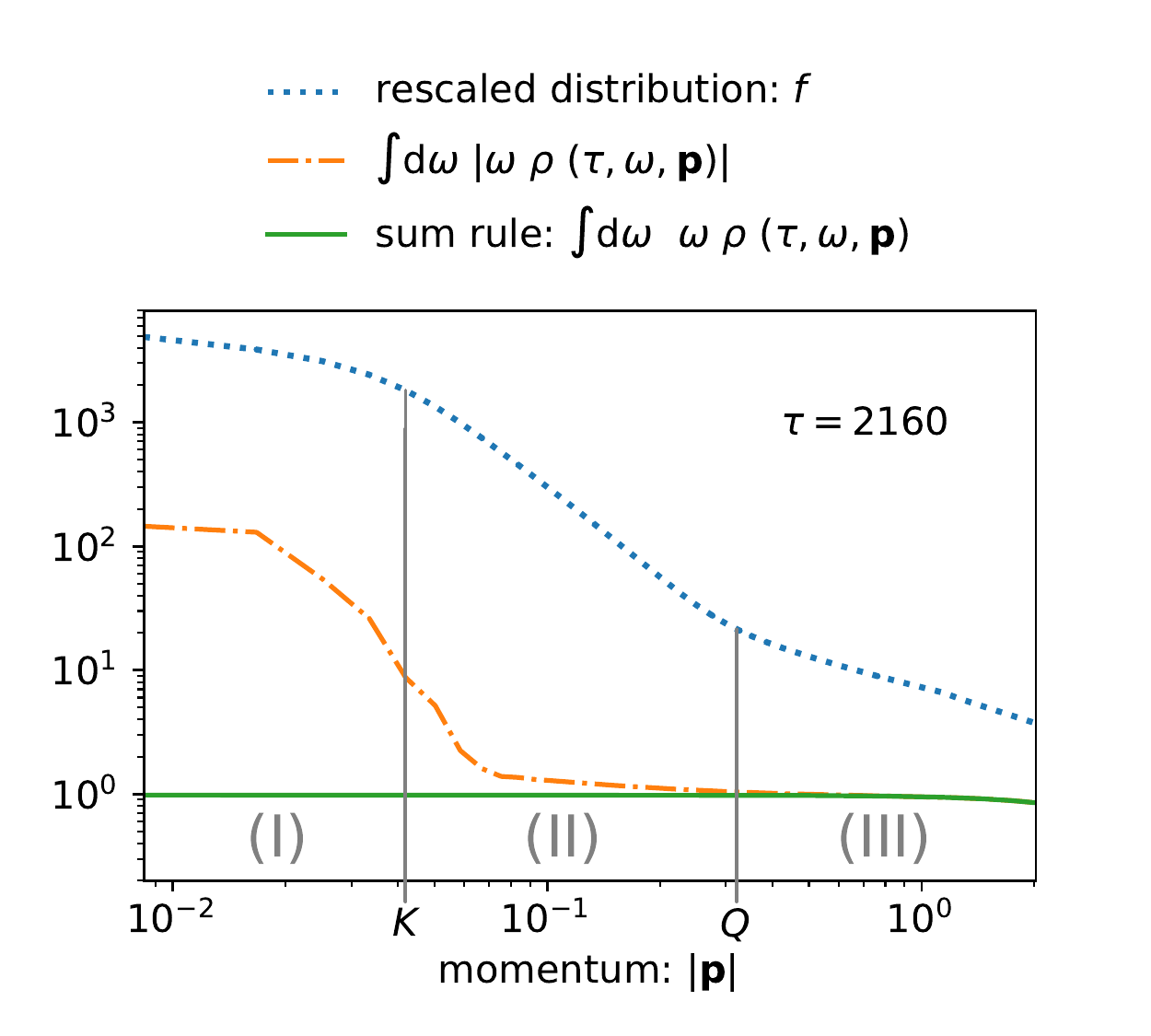}	
	\caption{Comparison of the sum rule for the spectral function (\ref{eq:sum_rule}) and the modified expression (\ref{eq:sum_rule_abs}) with the absolute value of the integrand as a function of momentum. 
	The significant deviations at low momenta result from negative values of $\rho (\tau, \omega, \mathbf{p})$. Shown is also the particle distribution function from which we identify the time-dependent scales $ K(t) $ and $ Q(t) $.}
	\label{fig:sum_rule}
\end{figure}

Using (\ref{eq:Wigner_transformation}), the momentum-space representation of the equal-time commutation relation \eqref{eq:commrel} can be written as the sum 
rule\footnote{The factor $ \omega $ in the integrand of \eqref{eq:sum_rule} appears due to the second-order time derivatives in the relativistic theory and is absent in the corresponding non-relativistic model.}
\begin{align}\label{eq:sum_rule}
\int _{0}^\infty \frac{\mathrm{d} \omega}{ \pi }
\ \omega\  \rho (\tau, \omega, \mathbf{p}) 
= 1. 
\end{align} 
We emphasize that in the interacting quantum field theory the spectral function $ \rho(\tau, \omega, |\mathbf{p}| ) $ is, in general, not positive for $ \omega > 0 $. In contrast, simple quasi-particle descriptions typically rely on positivity, such as the free-field spectral function with a positive particle peak at the frequency that is equal to the mass of the particle. We can check whether positivity is approximately realized by comparing to the corresponding expression with the absolute value of the integrand of \eqref{eq:sum_rule},   
\begin{align}\label{eq:sum_rule_abs}
\int _{0}^\infty \frac{\mathrm{d} \omega}{ \pi }\ |  \omega\  \rho (\tau, \omega, \mathbf{p}) |\,.
\end{align} 
Fig.~\ref{fig:sum_rule} compares this modified expression to the sum rule as a function of spatial momentum. Here the data for the Fourier transformed spectral function is computed using a discrete Fourier transformation, see Appendix~\ref{appendix:Wigner_Fourier_trafo} for details. For sufficiently large momenta, we indeed observe agreement as expected based on the validity of perturbation theory in this regime. However, at low momenta significant deviations are seen, indicating that there is no simple quasi-particle spectral function in the deep infrared.  

\begin{figure}
	\centering
	\includegraphics[width=0.5\textwidth, trim = 0 5 0 20]{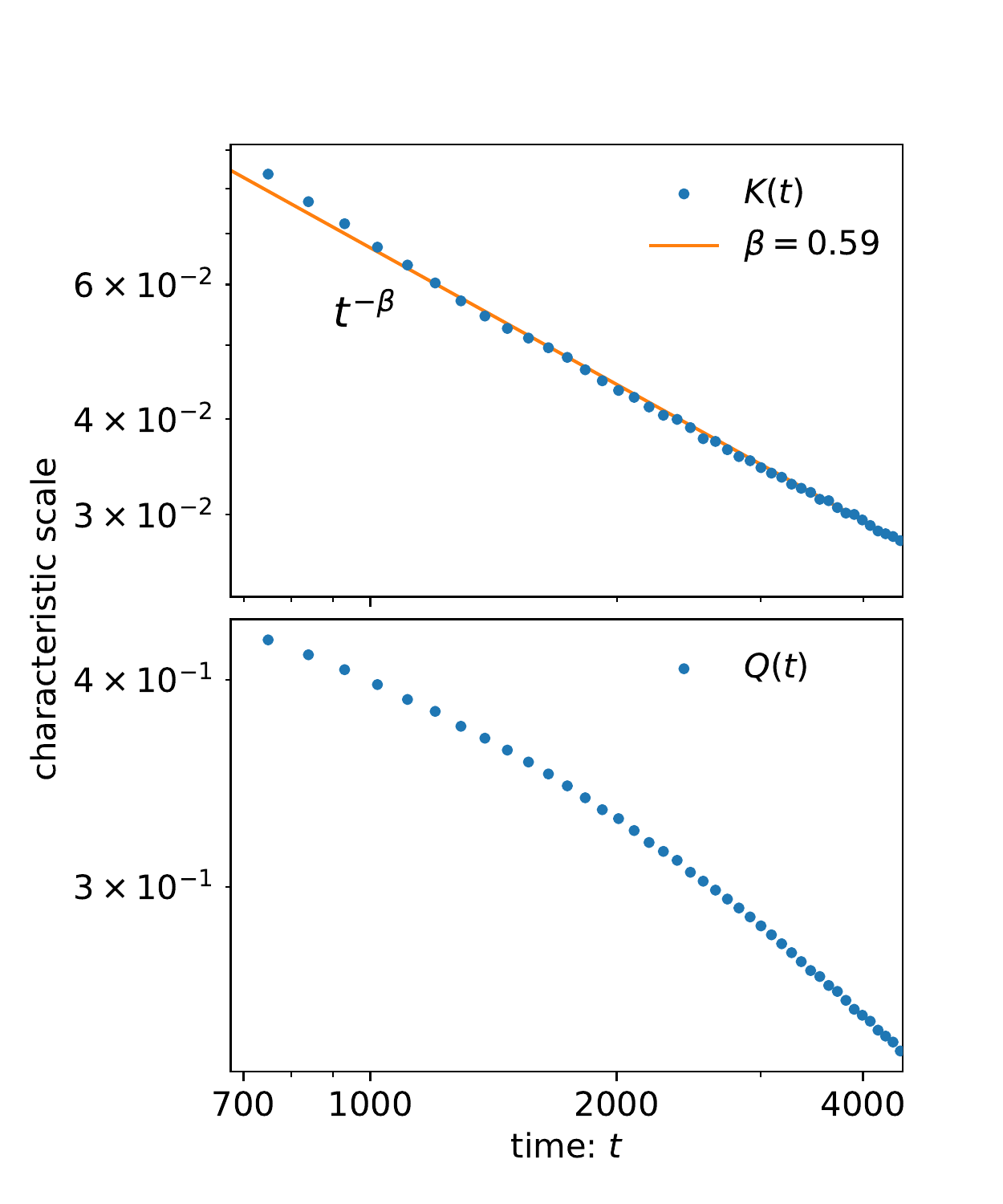}	
	\caption{
		Time evolution of the characteristic infrared momentum scale $ K(t) $ (upper plot) and the scale $ Q(t) $ separating the infrared and the high-momentum regime (lower plot). 
		The time evolution of the infrared scale becomes $ K(t) \sim t^{-\beta}$, while $ Q(t)$ evolves somewhat faster than a simple power-law. 
	}
	\label{fig:characteristic_momentum_scales}
\end{figure}

To further analyze this behavior, we distinguish three momentum regimes which we identify with the help of the particle distribution function defined in (\ref{eq:equaltime_particle_distribution}), as shown in Fig.~\ref{fig:sum_rule}. There we indicate a characteristic time-dependent momentum scale $K(t)$, where the distribution function shows maximum positive curvature in the scaling regime, along with $Q(t)$ defined by the scale of maximum negative curvature. These two scales are used to identify the 
\begin{enumerate}
	\item[(I)]
	infrared ``plateau'' regime, $ |\mathbf{p}| \ll K(t) $,
	\item[(II)]
	infrared ``power-law'' regime, $ K(t) \ll |\mathbf{p}| \ll Q(t) $,
	\item[(III)]
	high-momentum perturbative regime, $ |\mathbf{p}| \gg Q(t)$.
\end{enumerate}
We emphasize that both momentum regimes (I) and (II) constitute the inverse particle cascade and are characterized by the same universal scaling exponents. In particular, $ K(t) \sim t^{-\beta}$, which can be inferred from the upper graph of Fig.~\ref{fig:characteristic_momentum_scales}. The scale $ Q(t) $ also evolves towards the infrared as can be seen in lower plot of Fig.~\ref{fig:characteristic_momentum_scales}. It appears not to be described by a simple power law, as the evolution becomes faster with time.

In the following, we discuss the differences of the unequal-time two-point functions in the three momentum regimes in detail.

\subsection{Violations of the fluctuation-dissipation relation}
\label{section:fluct_diss_rel}

\begin{figure}
	\centering
	\includegraphics[width=.5\textwidth,
	 trim = 0 40 0 40
	]{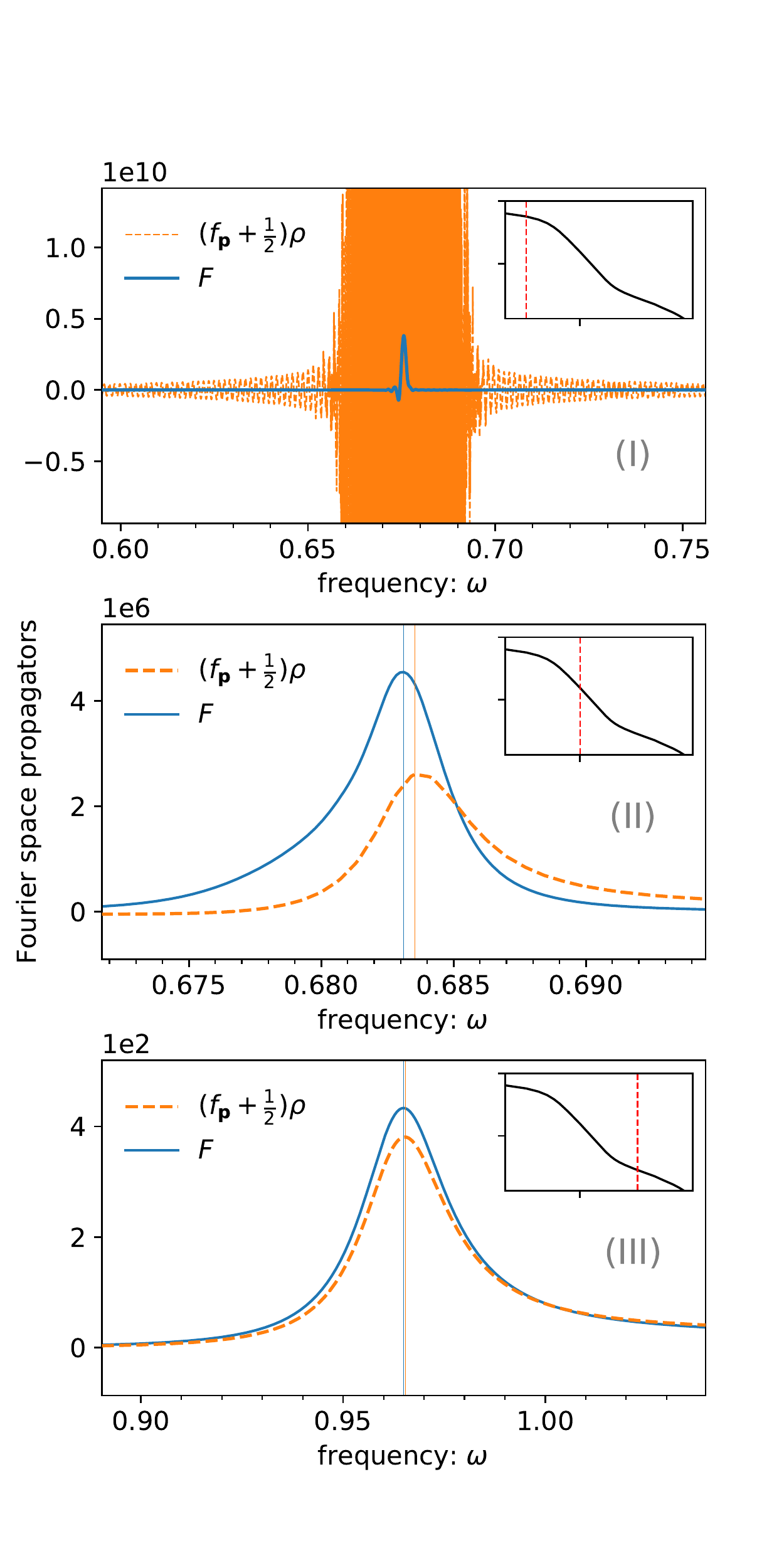}
	\caption{
		Comparison of $ \left(f(t=\tau, |\mathbf{p}|) +1/2\right) \rho(\tau, \omega, |\mathbf{p}|) $ and $ F(\tau, \omega, |\mathbf{p}|) $ as a function of frequency $ \omega $ at time $ \tau = 2250 $ for given momenta in the three regimes (I), (II) and (III) from top to bottom. 
		Here $ p_\mathrm{(I)} = 0.017 $, $ p_\mathrm{(II)} = 0.101 $ and $ p_\mathrm{(III)} = 0.679 $ are marked in the inset (red dashed line), where the distribution function $ f $ is plotted as a function of $ |\mathbf{p} |$ (black line) on a double logarithmic scale. 
	}
	\label{fig:Spectra_FW_3_regimes}
\end{figure}

In contrast to our non-equilibrium situation considered, thermal equilibrium is time-translation invariant, such that correlation functions only depend on relative coordinates and there is no $ \tau $-dependence. Moreover, the statistical and spectral functions in thermal equilibrium, $F^{(\mathrm{eq})}$ and $\rho^{(\mathrm{eq})}$, are related by the fluctuation-dissipation theorem, see e.g. \cite{Berges:2004yj}, according to
\begin{align}
F^{(\mathrm{eq})}(\omega, \mathbf{p}) = \Big(f_{\mathrm{BE}}(\omega) + \dfrac{1}{2}\Big)  \rho^{(\mathrm{eq})} (\omega, \mathbf{p}) \label{eq:FDR} \, .
\end{align}
This implies that the ratio $ F^{(\mathrm{eq})}(\omega, \mathbf{p})  /  \rho^{(\mathrm{eq})} (\omega, \mathbf{p})$ is independent of spatial momentum $ \mathbf{p} $ and determined by the frequency-dependent Bose-Einstein distribution $ f_{\mathrm{BE}} (\omega) = (e^{\beta\omega}-1)^{-1}$ with inverse temperature $\beta$ in the absence of conserved charges. 

Out of equilibrium, the statistical and spectral correlation functions are linearly independent in general, but one may hope to find some generalized fluctuation-dissipation relation where the Bose-Einstein distribution is replaced by a time-dependent distribution function. Such a generalized relation provides, for instance, the basis for standard kinetic descriptions. 

In order to study the non-equilibrium frequency space two-point functions, we consider the Fourier Wigner space propagators $ F(\tau, \omega, |\mathbf{p}|) $ and $ \rho(\tau, \omega, |\mathbf{p}|) $ for times $ \tau $ at which the system has reached the scaling regime. Details on the numerical computation of the frequency-space propagators can be found in Appendix~\ref{appendix:Wigner_Fourier_trafo}. 
As discussed above, we expect the spectral function to describe a quasi-particle excitation spectrum for sufficiently large momenta of regime (III) and maybe (II). 

Fig.~\ref{fig:Spectra_FW_3_regimes} shows the numerical results for $ (f(t=\tau, |\mathbf{p}|) +1/2) \rho(\tau, \omega, |\mathbf{p}|) $ and $ F(\tau, \omega, |\mathbf{p}|) $ as a function of $\omega$ at some time $\tau = 2250$ in the scaling regime for three different momenta in the ranges (I), (II), and (III) as indicated in the inset. 
As anticipated, from the upper graph one observes that both expressions are clearly different in the deep infrared regime (I). The middle graph shows that in regime (II) both quantities become similar in shape, however, the respective peaks are shifted relative to each other and the height is not the same. In contrast, $ F $ and $ (f + 1/2) \rho $ have the same Breit-Wigner shape for the considered momentum in regime (III) although the amplitudes do not fully agree yet. As the momenta become larger, we checked that this agreement gets more accurate. This establishes a well-defined generalized fluctuation-dissipation relation in terms of the non-equilibrium distribution function $ f(\tau, |\mathbf{p}|) $ in the high-momentum regime. 

\begin{figure}[t]
	\centering
	\includegraphics[width=.5\textwidth, 
	trim=0 30 0 30
	]{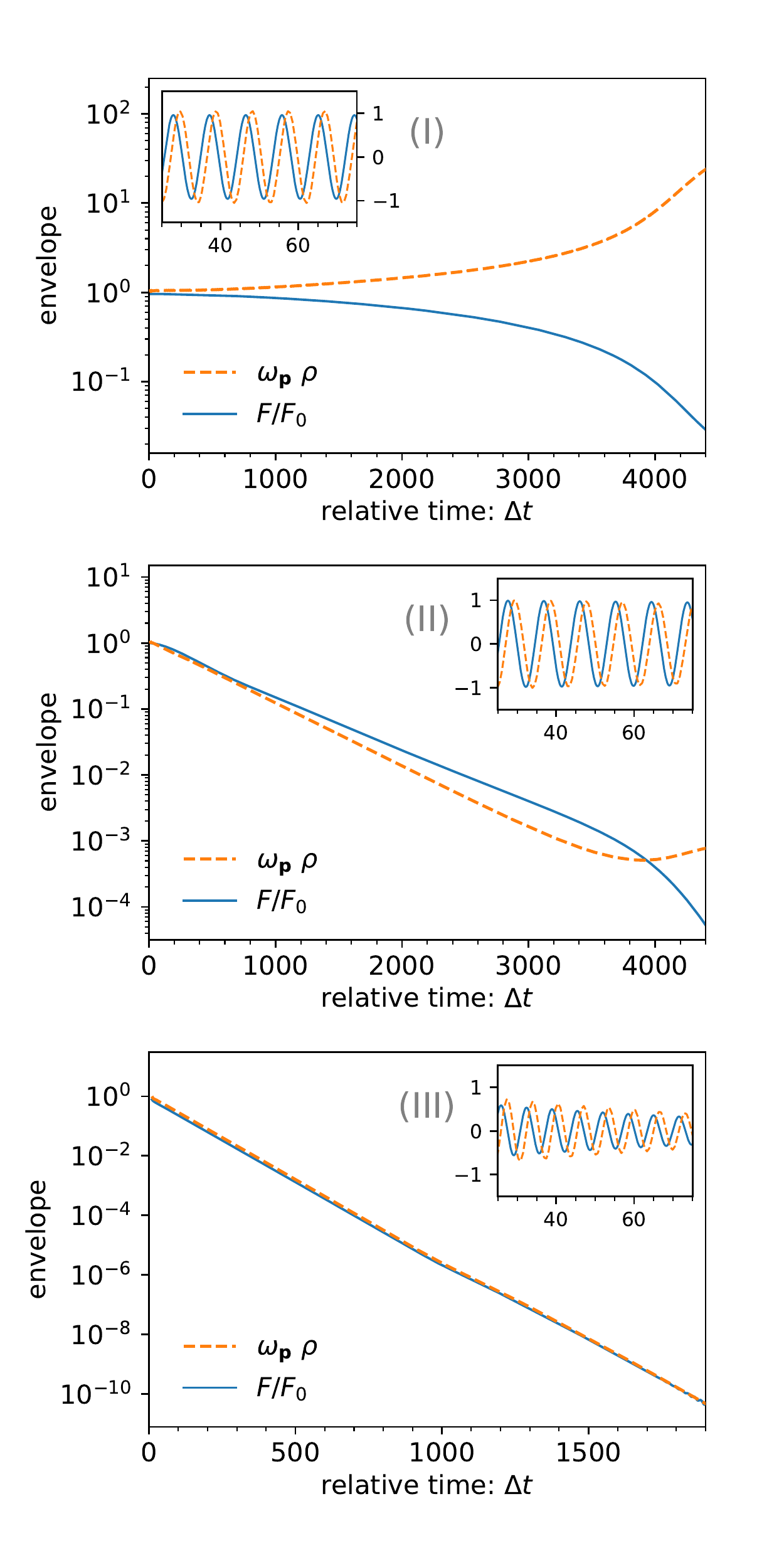}
	\caption{
		Envelopes of the spectral and statistical functions as a function of relative time $\Delta t$ at $ \tau = 2250 $, shown for the same momenta as in Fig.~\ref{fig:Spectra_FW_3_regimes}. 
		The spectral function is rescaled by the oscillation frequency $ \omega_\mathbf{p} $ and the statistical function by its maximum $ F_0 = F(\tau, \Delta t=0, |\mathbf{p}|) $
		The inset displays the oscillation in $ \Delta t $ at shorter times on a linear scale. 
		The time axis of the bottom plot shows a smaller time range since numerical uncertainties arise for envelopes becoming smaller than $ \mathcal{O}(10^{-9}) $. 
		}
	\label{fig:Spectra_W_3_regimes}
\end{figure}

We now analyze the behavior of the spectral function in regime (I) in more detail. 
The upper graph of Fig.~\ref{fig:sum_rule} reveals that the spectral function has the shape of an enveloped oscillation in this regime. 
We emphasize that this behavior is insensitive to the numerical discretization, i.e.\ to changes of the time-step size, the spatial grid step or volume size. Instead, it is related to the initial time boundaries at $ t\geq 0 $ and $ t'\geq 0 $, which enter the integration boundaries in \eqref{eq:Wigner_transformation}. 
In order to understand this effect better, it is helpful to consider the Wigner space propagators $ F(\tau, \Delta t, |\mathbf{p}|) $ and $ \rho(\tau, \Delta t, |\mathbf{p}|) $, i.e.\ without the Fourier transformation with respect to relative time $\Delta t$. Both functions oscillate in $ \Delta t $ with a frequency that in general can depend on both $ \tau $ and $ |\mathbf{p}| $. The envelopes of these oscillations are plotted in Fig.~\ref{fig:Spectra_W_3_regimes}, where the insets show the actual oscillations. The oscillation frequency can be used to define a dispersion relation $ \omega(\tau, |\mathbf{p}| ) $, which we call \textit{Wigner dispersion} in order to distinguish it from the effective dispersion defined in \eqref{eq:equaltime_dispersion}. Since the oscillation frequencies observed are practically constant during the self-similar time evolution, the Wigner dispersion is quasi-stationary.

In principle, one could obtain Wigner dispersions for the statistical and spectral functions separately. 
This seems necessary at first sight, since the behavior of $ F $ and $ \rho $ is clearly different in the infrared, as seen in Fig.~\ref{fig:Spectra_FW_3_regimes}. However, we find that
the relative difference $( \omega_F - \omega_\rho)/\omega_F$ is smaller than $ \SI{0.1}{\percent} $ such that for all practical purposes the Wigner dispersions of the statistical and spectral functions are treated as being equal. 

While the oscillation frequencies of $ F(\tau, \Delta t, |\mathbf{p}|) $ and $ \rho(\tau, \Delta t, |\mathbf{p}|) $ are practically equal, the envelopes of the oscillations can differ significantly from each other in the infrared. Fig.~\ref{fig:Spectra_W_3_regimes} compares the envelopes of $ F $ and $\rho$, which are plotted on a logarithmic scale against the relative time $ \Delta t $, in the different momentum regimes. In (II) and (III), one observes that both $ F $ and $ \rho $ decay exponentially in $ \Delta t $, which leads to well-defined quasi-particle peaks in frequency space as seen in Fig.~\ref{fig:Spectra_FW_3_regimes}. In regime (I), however, the spectral function grows towards the initial-time boundaries whereas the statistical function is damped stronger than exponentially (upper plot of Fig.~\ref{fig:Spectra_W_3_regimes}). 

To analyze this further, it is helpful to consider the simplified case of strictly exponentially damped oscillations. Omitting the $ \tau $-dependence for the moment, we approximately write
\begin{subequations}
	\begin{align}
	{F}(\Delta t, |\mathbf{p}| ) 
	&\simeq e^{- \gamma_\mathbf{p} |\Delta t|}
	\cos(\omega_\mathbf{p} \Delta t)\ F_0\,, \\
	{\rho}(\Delta t, |\mathbf{p}| ) 
	&\simeq e^{- \gamma_\mathbf{p}| \Delta t|}
	\sin(\omega_\mathbf{p} \Delta t) \ \omega_\mathbf{p}^{-1} \,,	\label{eq:exponentially_decaying_spectral_function}
	\end{align}
	\label{eq:exponentially_decaying_Wigner_fcts}%
\end{subequations}
where $ \gamma_\mathbf{p} $ is the damping constant, $ F_0 $ denotes the amplitude of the statistical function at $ \Delta t=0 $ and the factor $ \omega_\mathbf{p}^{-1}$ in \eqref{eq:exponentially_decaying_spectral_function} ensures that the spectral function suffices the commutation relation \eqref{eq:commrel}. For this \textit{ansatz}, calculating the Fourier transform with respect to the relative time $ \Delta t $ analytically yields the frequency space propagators
\begin{subequations}
	\begin{align}
F(\omega, |\mathbf{p}| ) 
&=F_0
\frac{
	2\gamma_\mathbf{p}  \ 
	\left (\omega^2 + \omega_\mathbf{p}^2\right)
}{( \omega^2 - \omega_\mathbf{p}^2)^2 + (2 \omega \gamma_\mathbf{p})^2},
\\
\rho(\omega, |\mathbf{p}| ) 
&=
\frac{4  \gamma_\mathbf{p}\  \omega}{( \omega^2 - \omega_\mathbf{p}^2)^2 + ( 2 \omega \gamma_\mathbf{p})^2},
\end{align}
	\label{eq:Breit_Wigner_fcts}%
\end{subequations}
where the latter corresponds to the relativistic Breit-Wigner function \cite{aarts2001nonequilibrium}. 
On-shell, where $ (\omega^2 + \omega_\mathbf{p}^2)  \approx 2 \omega^2 $, the statistical function is also described by a Breit-Wigner shape. 
The damping constant $ \gamma_\mathbf{p} $ determines the width of the quasi-particle peak. 

In regime (III), $F $ and $\rho $ decay with the same damping constant and consequently have the same Breit-Wigner shape in Fourier space, see Fig.~\ref{fig:Spectra_FW_3_regimes}. 
We can determine the parameters $ \gamma_\mathbf{p} $, $ \omega _\mathbf{p} $ and $ F_0 $ by fitting our data to either \eqref{eq:exponentially_decaying_Wigner_fcts} or \eqref{eq:Breit_Wigner_fcts}. Since the Fourier Wigner space propagators are numerically obtained by a discrete Fourier transformation of the Wigner propagators, both methods are equivalent up to numerical uncertainties.

\begin{figure*}[t]
	\centering
	\includegraphics[width=1.0\textwidth, trim=0 25 0 15]{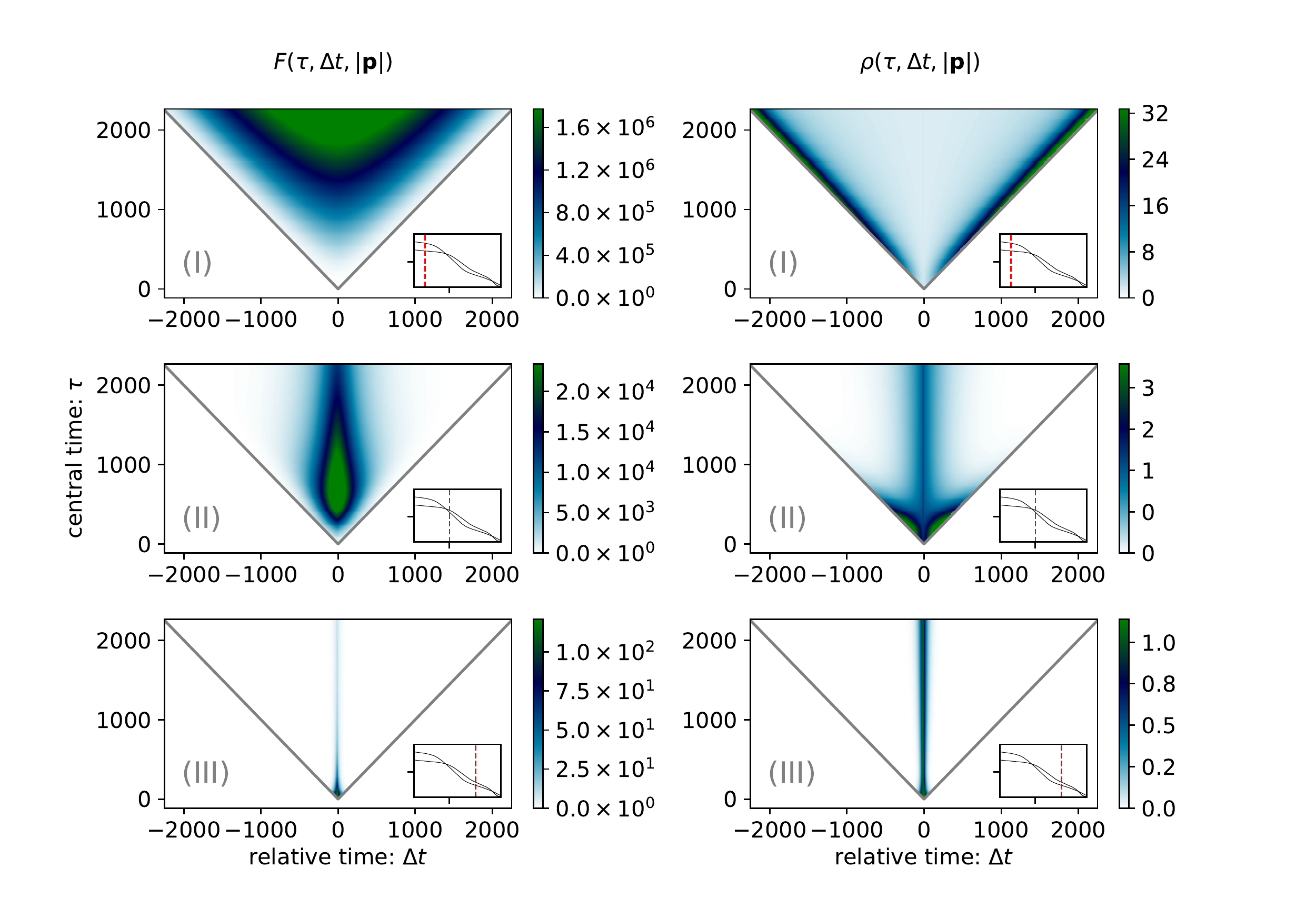}
	\caption{
		The spectral and statistical functions as a function of central time $\tau$ and relative time $\Delta \tau$ for the three different momenta as indicated by the dashed line in the insets. The latter show the distribution functions as a function of momentum for two different times.
	}
	\label{fig:contour_F_and_R}
\end{figure*}

When looking at regime (II) the statistical function has a slightly larger damping constant than the spectral function. The main difference, however, appears at large relative times $ \Delta t $, i.e.\ at the temporal boundaries $ t=0 $ and $ t'=0 $, where the spectral function grows somewhat while the statistical function decays even faster. This means that spectral correlations at large time-separations are enhanced whereas statistical correlations are suppressed. As a consequence, the Fourier Wigner propagators shown in Fig.~\ref{fig:Spectra_FW_3_regimes} reveal different shapes. 

For exponentially decaying Wigner space propagators, the effects from finite integration bounds are negligible at sufficiently late times, where $ \tau $ is large. However, in regime (I) the spectral function does not decay towards the initial-time boundaries whereas the statistical function drops very quickly. Consequently, only the statistical function is described by a peak in Fourier Wigner space, however, the peak cannot be described by a Breit-Wigner function since no exponential decay is involved. 
In contrast, the growth of the spectral function at large relative times is sharply cut off by the time boundary $ |\Delta t| < 2 \tau $ appearing in initial value problems. 
Hence, the Wigner transformation gives rise to fast oscillations of the spectral propagator in frequency space, as seen in the upper plot of Fig.~\ref{fig:Spectra_FW_3_regimes}.
The oscillation frequency is determined by the integration range. The spectral propagator $ \rho(\tau, \omega, |\mathbf{p}|) $ oscillates in $ \omega $ with frequency $ 2 \tau $ and has an envelope that is peaked around $ \omega = \omega_\mathbf{p} $. 

The differences of the Wigner space propagators between the three momentum regimes can be visualized more clearly when looking at the contour plots shown in Fig.~\ref{fig:contour_F_and_R},
where the relative time $ \Delta t $ labels the horizontal axis, the central time $ \tau $ the vertical axis, and the envelopes of $F(\tau, \Delta t, |\mathbf{p}|) $ and $ \rho(\tau, \Delta t, |\mathbf{p}|) $ are encoded in the color scheme. 
The initial time bounds $ t, t' \geq 0 $, equivalent to $ - 2 \tau \leq \Delta t \leq 2 \tau  $, are marked by the gray lines.
Each horizontal slice corresponds to a fixed central $ \tau $ and can be Wigner-transformed with respect to $ \Delta t$ in order to obtain the corresponding Fourier Wigner space propagators. The propagators shown in Fig.~\ref{fig:Spectra_FW_3_regimes} correspond to the latest available time slices at $ \tau = 2250 $. 

In accordance with the behavior of the particle distribution function discussed above, the amplitude of $ F $ decreases by several orders of magnitude when going from low to high momenta (left plots of Fig.~\ref{fig:contour_F_and_R} from top to bottom).
In contrast, the spectral function $ \rho $ does not differ much in amplitude since it is normalized according to the sum rule \eqref{eq:sum_rule}. 

The contour plots in Fig.~\ref{fig:contour_F_and_R} visualize how the envelopes of $ F $ and $ \rho $ evolve with time $ \tau $. Due to the self-similar time evolution, going towards later times $ \tau $ is equivalent to moving to larger momenta $ |\mathbf{p}| $. In the infrared momentum regime, the  increase of the amplitude of the spectral function towards the initial time bounds declines with evolving time $ \tau $. 
For higher momenta, the decay rate increases as the time $ \tau $ evolves. Hence, the effect of high amplitudes at large $ \Delta t$ vanishes at sufficiently late times, where the time scale is larger for small momenta. If one considers, for instance, some fixed momentum $|\mathbf{p}^*| < K(t) $ in region (I), then during the time evolution the characteristic momentum $ K(t) ~ \sim t^{-\beta}$ moves towards the infrared such that at some later time $t' > t$ one finds $ |\mathbf{p}^*|  > K(t') $. Since the momentum moved from region (I) into region (II), where the Wigner space propagators decay exponentially, no boundary effects occur. In that sense, due to the self-similar time-evolution it is always possible to wait long enough to overcome the initial-time boundary effect for a given momentum.

\subsection{Dispersion and effective mass}
\label{sec:effmass}

\begin{figure}[t] 
\centering
\includegraphics[width=.5\textwidth, trim = 0 10 0 30]{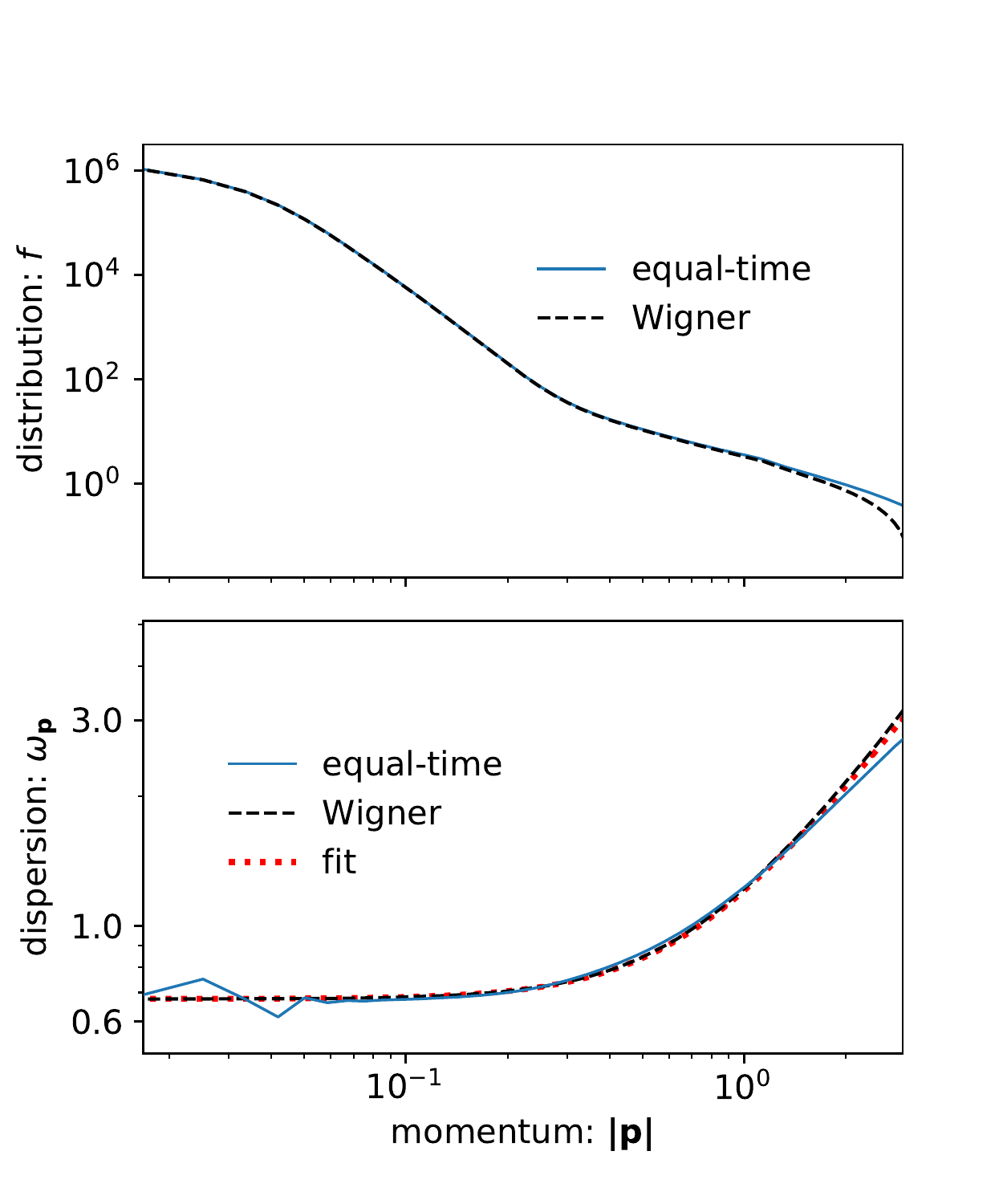}	
\caption{
	The distribution function $ f $ and the dispersion relation $ \omega $ at time $ t = \tau = 2250 $. The quasi-particle definition and the Wigner space definition agree very well. The Wigner dispersion is fitted to the relativistic dispersion relation $ \sqrt{|\mathbf{p}|^2 + m_\mathrm{eff}^2} $ with $ m_\mathrm{eff} \approx \SI{0.68}{} $. 
}
\label{fig:Comp_equal_unequal_time}
\end{figure}

Motivated by the results of the last section, we may consider a ``Wigner'' particle distribution function $f( \tau , |\mathbf{p}| )$ obtained from  
\begin{align}
f( \tau , |\mathbf{p}| ) + \dfrac{1}{2} = \int_0 ^ \infty \dfrac{\mathrm{d} \omega}{\pi }\  \omega\  F(\tau, \omega,| \mathbf{p}|), 
\label{eq:unequaltime_particle_distribution}
\end{align}
which is a priori different from the ``equal-time'' definition employed in (\ref{eq:equaltime_particle_distribution}). From the upper panel of Fig.~\ref{fig:Comp_equal_unequal_time} one observes that both definitions are in good agreement with each other. Small deviations in the high-momentum range can be cured in the limit $ a_t \rightarrow 0 $. 

Similarly, we introduced the effective dispersion \eqref{eq:equaltime_dispersion} and the Wigner dispersion (see Sect\ \ref{section:fluct_diss_rel}). Both definitions for the dispersion relations can actually be fitted rather well to a relativistic dispersion, 
\begin{align}
\omega(|\mathbf{p}| ) = \sqrt{|\mathbf{p}|^2 + m_{\mathrm{eff}}^2},
\label{eq:rel_dispersion}
\end{align}
with effective mass $  m_\mathrm{eff}$ that incorporates quantum-statistical fluctuations. Our numerical computations show that they agree well with each other, as seen from the lower graph of Fig.~\ref{fig:Comp_equal_unequal_time}. 
Although both the dispersion and the effective mass are in general time-dependent, they turn out to be practically constant in time in the scaling regime. 
Because the extraction of the effective mass from the Wigner dispersion turns out to be numerically more stable than using the equal-time dispersion, the values cited in the text 
are obtained from the former. In particular, we find from 
fitting the data to the relativistic dispersion relation (\ref{eq:rel_dispersion}) the following values for the effective mass for different values of the interaction parameter, 
\begin{align}
m_\mathrm{eff}  = 
\begin{cases}
\SI{0.638(5)}{}& \lambda = 0.01\\
\SI{0.641(5)}{}& \lambda = 0.10 \\
\SI{0.677(7)}{}& \lambda = 1.00 \\
\SI{0.716(8)}{}& \lambda = 1.00 
\end{cases}
\end{align} 
where the error indicates that $ m_\mathrm{eff} $ is not exactly time-independent during the self-similar evolution. 
The presence of an effective mass explains why the infrared exponents found in Sec.~\ref{section:self-similar_dynamics} are close to the predictions for a non-relativistic theory~\cite{Orioli:2015dxa}. 
As can be seen in the upper plot of Fig.~\ref{fig:Occupation_Original_and_Rescaled}, the momenta of the whole infrared regime are much smaller than the effective mass (note the log-scale) 
and thus effectively non-relativistic. 

\section{Conclusion}
\label{sec:conclusion}

We analyzed the far-from-equilibrium scaling properties of equal-time and unequal-time correlation functions in scalar quantum field theory. The numerical results are obtained from a fully self-consistent large-$N$ expansion to NLO. Our results for scaling exponents and scaling functions are in agreement within errors with previous weak-coupling estimates for equal-time correlations using effective kinetic theory or classical-statistical field theory. We find these universal results for a wide range of couplings even beyond the weak-coupling regime. Moreover, we have established the self-similar behavior of the strongly momentum-dependent effective coupling.  

The computation of the unequal-time spectral and statistical functions allowed us to observe the validity of a generalized fluctuation-dissipation relation in the perturbative regime at high
momenta, while we demonstrated that significant violations occur in the non-perturbative infrared. We identified a characteristic deep-infrared regime, where the corresponding distribution function approaches a ``plateau''. In this regime the spectral function does not decay as a function of relative time, which leads to an enhanced sensitivity to initial times and the absence of positivity for 
frequencies $\omega > 0$. The statistical function in this regime shows a characteristic peak structure with significant deviations from a Breit-Wigner form.  

Our results give unprecedented insights into the non-perturbative nature of collective excitation far from equilibrium. While many features of the system are indeed seen to become universal near 
a non-thermal fixed point also beyond the weak-coupling regime, unequal-time properties encoded in spectral functions can reveal intriguing properties such as an enhanced sensitivity to initial times. Since our results are based on a large-$N$ expansion, it would be very interesting to also analyze the behavior of weakly-coupled systems with small $N$ using real-time lattice simulations along the 
lines of Refs.~\cite{PineiroOrioli:2018hst,Boguslavski:2019ecc}.

\section*{Acknowledgments}

We thank Kirill Boguslavski, Asier Pineiro Orioli and Alexander Rothkopf for discussions and collaborations on related work. 

The authors acknowledge support by the state of Baden-W\"urttemberg through bwHPC. This work is part of and has been supported by the DFG Collaborative Research Centre ``SFB 1225 (ISOQUANT)".

\appendix

\section{Extracting scaling exponents of the self-similar time evolution}
\label{appendix:scaling_exponents_method}

In this appendix, we present the fitting procedure employed to determine the scaling exponents $ \alpha $ and $ \beta $ from the numerical data. Our method is similar to the approach put forward in \cite{Orioli:2015dxa} and \cite{Karl:2016wko}. 

Numerically, we approximate the scaling function in the universal scaling regime by the rescaled function,
\begin{align}
f_\mathrm{resc} (t, \mathbf{p}) = \left( \tfrac{t}{ t_\mathrm{ref}} \right) ^{-\alpha} f \left ( t,  \left( \tfrac{t}{ t_\mathrm{ref}} \right )^{-\beta} \mathbf{p} \right ),
\end{align}
which allows us to compare the distribution at a reference time $ t_\mathrm{ref} $ with several distributions at earlier times $ t_i $, $ i = 1, \ldots, n $, within a given time window $ \Delta t_\mathrm{w} $. For the correct set of scaling exponents, a perfect self-similar time evolution implies that the rescaled distribution becomes independent of time, i.e. 
\begin{align}
f_\mathrm{resc}(t_\mathrm{ref}, \mathbf{p}) =  f_\mathrm{resc}(t_\mathrm{i}, \mathbf{p}) 
\end{align}
for times $ t_i $ and $ t_\mathrm{ref} $ within the scaling regime.
 
In practice, the numerically obtained distribution functions deviate from the perfect scaling behavior. The scaling exponents are determined by minimizing these deviations, which we quantify by 
\begin{align}	
\chi ^2(\alpha, \beta) 
= \dfrac{1}{n} \sum_{i=0}^{n}
\int \d\, \log \mathbf{p} 
\left[
\Delta _i (\mathbf{p})
\right] ^2
\label{eq:chi_sqrt}
\end{align}
with
\begin{align}
\Delta_i (\mathbf{p})=  \log f_\mathrm{resc}(t_\mathrm{ref}, \mathbf{p}) - \log  f_\mathrm{resc}(t_i, \mathbf{p}) \,.
\end{align}
We sum over the $ n $ comparisons between the reference time $ t_\mathrm{ref} $ and earlier times $ t_i < t_\mathrm{ref}$.  The integration over $ \mathrm{d} \log \mathbf{p} $ enhances the low-momentum range. 

In order to compute the differences $ \Delta_i (\mathbf{p})$, the numerical data of the distribution functions is interpolated using a cubic spline provided by the Python SciPy library. The integration range is chosen dynamically in terms of the time-dependent characteristic momentum scales $ K(t) $ and $ Q(t) $. We include $ \SI{95}{\percent} $ of the momentum range $ [\log \Lambda_\mathrm{IR}, \log K(t_\mathrm{ref})]  $ and $ \SI{92}{\percent} $ of the momentum range $ \left [\log K(t_\mathrm{ref}), \log Q(t_\mathrm{ref})  \right ]$, which ensures that the high-momentum range is excluded from the fitting procedure for the analysis of the infrared fixed point. 

From the minimization of \eqref{eq:chi_sqrt} at different reference times we obtain scaling exponents $ \bar{\alpha}$ and $\bar{ \beta} $ at which $ \chi^2 $ is minimal. 
Here, we use optimization routines of Python SciPy library for the minimization procedure. 
Thereby we observe that $  \chi ^2 (\bar{\alpha}, \bar{ \beta}) $ decreases with time and converges to a constant value when the system is approaching the non-thermal fixed point. 

The error for the values of $ \bar{\alpha}$ and $\bar{ \beta} $ is estimated by the marginal likelihood functions  
\begin{align}
W_\beta(\alpha) &= \mathcal{N}_\alpha \exp \left[ - \dfrac{\chi ^2 (\alpha, \bar{\beta})}{2 \chi ^2(\bar{\alpha}, \bar{\beta})}\right], \\
W_\alpha(\beta) &= \mathcal{N}_\beta \exp \left[ - \dfrac{\chi ^2 (\bar{\alpha}, {\beta})}{2 \chi ^2(\bar{\alpha}, \bar{\beta})}\right], 
\end{align}
where the normalization constants $ \mathcal{N}_{\alpha,\beta} $ are chosen such that $ \int  \mathrm{d} \alpha \ W(\alpha) = \int \mathrm{d} \beta \  W(\beta) = 1$. Fitting $ W _{\alpha,\beta}$ to a Gaussian distribution allows us to estimate the error of the exponents by the standard deviation of the Gaussian function. We refer to this uncertainty as the \textit{fit errors} denoted by $ \Delta \alpha_\mathrm{fit} $ and $ \Delta \beta _\mathrm{fit} $, which determine the errorbars in the plots shown in Figs.~\ref{fig:scaling_exponents} and \ref{fig:scaling_exponents_a_lambda}. In the analysis of the distribution function $ f$, the fit errors of both exponents $ \alpha $ and $ \beta $ decrease with the time evolution and approach asymptotic values. 

The asymptotic values of the exponents are obtained by averaging over the values obtained at late reference times. For the analysis of the main text we compute the mean over a time window $ \Delta t_\mathrm{av} \geq 600$. The corresponding standard deviation is used to quantify the \textit{statistical errors} $ \Delta \alpha_\mathrm{stat} $ and $ \Delta \beta_ \mathrm{stat} $. These errors reflect how strong the exponents fluctuate during the time-evolution. 
For the values presented in Tables~\ref{tab:exponent_values} and \ref{tab:exponent_values_l_eff}, we provide the error
\begin{align}
\Delta \alpha = \sqrt{(\Delta \alpha_\mathrm{fit})^2 + (\Delta \alpha_\mathrm{stat})^2 },
\end{align}
and accordingly for the other exponents. 

In the scaling analysis presented, $ n=4 $ comparisons within a time window of $ \Delta t_\mathrm{fit} = 720 $ were used. The fit and statistical errors of our simulations are shown in Tables~\ref{tab:exponent_error_values} and~\ref{tab:exponent_error_values_l_eff}. While the analysis for $ f $ yields very small statistical errors, the exponents obtained for both the effective coupling $ \lambda_\mathrm{eff} $ as well as the mode energy $ \varepsilon $ reveal strong fluctuations represented by much larger statistical errors.

We have checked that the values for the exponents are insensitive to the numerical discretization (time step size and cutoffs). 
Naturally, there exists a dependence on the parameters of the method used, such as the number of comparisons $ n $, the time windows $ \Delta t_\mathrm{w} $ and $ \Delta t_\mathrm{fit} $, and the momentum range used to compute \eqref{eq:chi_sqrt}. We checked that the values obtained are not sensitive to $ n $ and $ \Delta t _{w} $. 

\begin{table}[h]
	\centering
	\begin{ruledtabular}
		\begin{tabular}{ccccccc}
			&$\lambda$& $\Delta \alpha_\mathrm{fit}$ &$\Delta \alpha_\mathrm{stat}$
			& $\Delta\beta_\mathrm{fit}$	& $\Delta\beta_\mathrm{stat}$&\\
			\hline
&$\SI{0.01}{} $ 
&$ \SI{0.06}{}$& $ \SI{0.02}{}$ &$\SI{0.02}{}$&$\SI{0.001}{}$&\\
&$\SI{0.10}{} $ 
&$ \SI{0.06}{}$& $\SI{0.02}{}$ & $\SI{0.02}{}$&$\SI{0.004}{}$&\\
&$\SI{1.00}{} $ 
&$ \SI{0.08}{}$& $\SI{0.02}{}$ &$\SI{0.02}{}$&$ \SI{0.004}{}$&\\
&$\SI{2.00}{} $ 
&$ \SI{0.09}{}$& $\SI{0.02}{}$ &$\SI{0.02}{}$&$ \SI{0.004}{}$&\\
	\end{tabular}
	\end{ruledtabular}
	\caption{Fit errors and numerical errors for the scaling analysis of $ f $ presented in the main text. The methods are described in Appendix \ref{appendix:scaling_exponents_method}.}
	\label{tab:exponent_error_values}
\end{table}

\begin{table}[h]
	\centering
	\begin{ruledtabular}
		\begin{tabular}{ccccccc}
			&$\lambda$& $\Delta \alpha_{\lambda,\mathrm{fit}}$ &$\Delta \alpha_{\lambda,\mathrm{stat}}$
			& $\Delta\beta_\mathrm{fit}$	& $\Delta\beta_\mathrm{stat}$&\\
			\hline
&$\SI{0.01}{} $ 
&$ \SI{0.22}{}$& $ \SI{0.28}{}$ &$\SI{0.08}{}$&$\SI{0.10}{}$&\\
&$\SI{1.00}{} $ 
&$ \SI{0.28}{}$& $\SI{0.25}{}$ &$\SI{0.10}{}$&$ \SI{0.09}{}$&\\
		\end{tabular}
	\end{ruledtabular}
	\caption{Fit errors and numerical errors for the scaling analysis of $ \lambda_\mathrm{eff} $ presented in the main text. The methods are described in Appendix \ref{appendix:scaling_exponents_method}.}
	\label{tab:exponent_error_values_l_eff}
\end{table}

\section{Wigner transformation}
\label{appendix:Wigner_Fourier_trafo}

In this appendix we present the methods used in order to compute the Wigner transformed spectral and statistical functions according to \eqref{eq:Wigner_transformation}.
In order to compute the Wigner transform $ f(\omega) $ of a temporal signal $ f(t) $, we need to numerically evaluate integrals of the form
\begin{align}\label{eq:continuous_FT}
f (\omega)
&= 
\int  _{-2\tau}^{2\tau} \mathrm{d}t \ 
e^{i\omega t} 
f(t)\,,
\end{align} 
where we use $ t $ instead of $ \Delta t $ here for notational convenience. 
We are interested in signals oscillating with a given frequency $ \nu $ that are enveloped by some function $ g(t) $ which can be written as
\begin{align}\label{eq:signal}
f_{\pm}(t) = g(t) \ \frac{1}{2}( e^{i\nu t} \pm e^{-i\nu t})\,,
\end{align}
where the relative sign determines whether the signal is symmetric or antisymmetric. 
In the following, we present the methods that we employ in order to compute \eqref{eq:continuous_FT} for such signals. \\

If $ g(t) $ decays sufficiently strong, i.e.\ if it becomes sufficiently small at the boundaries $ \pm 2 \tau $, effects from the finite integration boundaries are negligible. In this case, the Wigner transform can be computed using a discrete Fourier transformation (DFT), 
\begin{align}
f(\omega_m)
=
\sum_{n=0}^{N_t-1}	f(t_n) e^{i 2\pi  m  n/N_t  }  \,,
\end{align}
where $ N_t $ denotes the number of data points, and $ \omega_m$ and $ t_m $ are the discretized frequency and time, respectively, with $ m = 0, \dots, N_t - 1 $. 
In our simulations, the DFT is implemented using the FFTW library \cite{fftw}. 

For an exponentially decaying envelope, which is described by $ g(t) \sim \exp( - \gamma | t|) $ with some decay constant $ \gamma $, the DFT can be compared to the analytically calculated Wigner transform with finite as well as infinite integration boundaries. 
For the parameters relevant in our simulations, we confirmed that the DFT of the exponentially decaying Wigner functions appearing in the high momentum regime yields accurate results. 
In addition, we checked the applicability of the DFT for other momentum ranges by comparing the decaying behavior of the propagator functions with exponential decays. 
Thus, for the analysis shown in the main text we use the DFT to compute the Wigner transform of the statistical function in all momentum ranges and the spectral function for medium or high momenta. \\

In general, the finite integration boundaries in \eqref{eq:continuous_FT} lead to oscillations of the Wigner transform $ f(\omega) $. These are particularly relevant if the envelope function $ g(t) $ increases towards larger $| t| $.
In this case, we determine the envelope $ g(t) $ by a polynomial fit and the oscillation frequency $ \nu $ using the Lomb-Scargle periodogram implemented in the Python SciPy library. We can then analytically compute the Wigner transformation of \eqref{eq:signal} for the relevant parameters. 
This method is employed for computing the spectral function at small momenta. However, when comparing the Wigner transform at different spatial momenta $ |\mathbf{p}| $ as in Fig.~\ref{fig:sum_rule}, we employ the DFT for both $ F $ and $ \rho $. 

\section{Computation of the time evolution}

The time evolution is computed numerically using the Euler discretization scheme for the temporal derivatives. 
Since the evolution equations involve second order time derivatives, specifying initial conditions at the numerical time steps $ t=0 $ and $ t=1 $ allows us to compute the time evolution iteratively for numerical time steps $ t \geq 2 $. The initial conditions used are
\begin{subequations}
	\begin{align}
	F(0, 0, |\mathbf{p}|) &= 	\dfrac{f_\mathbf{p} + \tfrac{1}{2}}{\omega_\mathbf{p}} ,\\
	F(1,0, |\mathbf{p}|) &= F(0, 0, |\mathbf{p}|),
	\label{eq:init_cond_GF}\\ 
	F(1,1,|\mathbf{p}|) &= F(0, 0, |\mathbf{p}|) \Big( a_t^2 \omega_\mathbf{p} ^2 + 1 \Big)\,,
	\end{align}
	\label{eq:boson_initF}%
\end{subequations}
with $ f_\mathbf{p} $ specified by the box initial conditions \eqref{eq:initial_distribution} and $ \omega_{\mathbf{p}} = \sqrt{\mathbf{p}^2 + m_0^2} $ in the limit $ m_0^2 \rightarrow 0 ^+$. \\

The non-equilibrium time evolution is computed from the evolution equations of the form
\begin{align}
\left[
\partial_t^2 + \mathbf{p}^2 + M^2(t)
\right]
F(t, t', |\mathbf{p}|)
&= I_\text{memory}\,,
\end{align}
see \cite{Berges:2016nru} for a summary of the relevant expressions. 
Here, $ I_\text{memory} $ denotes the causal memory integrals and the effective mass squared is given by
\begin{align}
M^2(t) &= m^2 
+ \dfrac{\lambda(N+2)}{6N} \int_\mathbf{p}
F(t, t, |\mathbf{p}|) \,,
 \nonumber
\end{align}
with $ m\rightarrow 0 $. In this limit, the tadpole contribution at initial time is determined by
\begin{align}
\int_\mathbf{p}
F(0, 0, |\mathbf{p}|) 
&=  
\int \mathrm{d} |\mathbf{p}| \, |\mathbf{p}|  \, 
	\left( f_\mathbf{p} + \tfrac{1}{2}\right )
\end{align}
with an ultraviolet cutoff given by the discretization. 

\bibliography{references}

\end{document}